\documentclass[
 reprint,
superscriptaddress,
 amsmath,amssymb,
 aps,
]{revtex4-2}

\usepackage{graphicx}
\usepackage{dcolumn}
\usepackage{bm}
\usepackage{hyperref}
\usepackage{xcolor}
\usepackage[T1]{fontenc}
\usepackage[utf8]{inputenc}
\usepackage{siunitx}
\usepackage{caption}
\usepackage{subcaption}
\usepackage{amsmath}
\usepackage{braket}
\usepackage{ragged2e} 
\DeclareCaptionJustification{justified}{\justifying}
\hypersetup{colorlinks,linkcolor={red!50!black},citecolor={blue!50!black},urlcolor={blue!80!black}}
\usepackage{pgf}
\usepackage{pgfplots}
\usepackage{mathrsfs}

\newcommand{\affA}{Institute for Molecular Science, National Institutes of Natural Sciences, Okazaki, Japan}
\newcommand{\affB}{RIKEN Center for Quantum Computing (RQC), 351-0198 Wako, Japan}
\newcommand{\affC}{Institute for Quantum Electronics, ETH Zürich, Otto-Stern-Weg 1, 8093 Zürich, Switzerland}
\newcommand{\affD}{Quantum Center, ETH Zurich, CH-8093, Switzerland}

\begin{document}

\title{Protocols of coherent motion control for an interaction-driven Rydberg gate}

\author{Valentin Magro}
  \email{valentin-magro@ims.ac.jp}
  \thanks{These authors contributed equally.}
  \affiliation{\affA}
\author{Wojciech Adamczyk}
  \email{wadamczyk@phys.ethz.ch}
  \thanks{These authors contributed equally.}
  \affiliation{\affC}\affiliation{\affD}
\author{Sylvain de L\'es\'eleuc}
  \email{sylvain.deleseleuc@riken.jp}
  \affiliation{\affB}

\date{\today}

\begin{abstract}

Generating entanglement between two Rydberg atoms is at the core of neutral-atom quantum computers. Current two-qubit gates operate in the Rydberg-blockade regime, in which the full strength of the van der Waals interaction between the two Rydberg atoms is not directly exploited, to avoid sensitivity to the position noise of the tweezer-trapped atoms, at the cost of a longer time spent in the Rydberg state. Here, we propose a set of techniques based on coherent control of the atomic motion obtained by combining optical tweezers and a two-dimensional optical lattice, and a sequence of multiple on/off pulses. The protocols keep the two-qubit gate error contribution from position noise below $10^{-4}$, heat the atom by less than~$\Delta n = 0.01$, while being robust to alignment errors of the potential up to $50$~nm and thermal excitation up to $\bar{n} = 3$. This toolbox opens the path for new two-qubit Rydberg gates directly, or partially, driven by the interaction, in which the atoms spend only $\sim 10$~ns in the Rydberg state, minimizing the increasingly dominant error source originating from its finite lifetime.

\end{abstract}

\maketitle

\section{Introduction}

The overhead of fault-tolerant quantum computing depends very sensitively on the physical two-qubit error rate \cite{Fowler2012Surface}, so pushing gate fidelities higher remains a central challenge for all hardware platforms. In neutral-atom quantum processors, two-qubit entangling gates are commonly implemented using the strong interactions between atoms excited to Rydberg orbitals. The seminal proposal for Rydberg-mediated entanglement introduced two distinct gate mechanisms: the interaction-driven gate and the Rydberg-blockade gate \cite{Jaksch2000Fast}. Experimental progress has so far been dominated by the blockade gate, whose fidelity has now reached the range of 99.5 to 99.9\% \cite{Evered_2023, Lib2026Velocity, Radnaev2025Universal, Tsai2025Benchmarking, Muniz2025High, Senoo2025High, evered2026high} through continued suppression of technical noise sources, in particular electric-field fluctuations affecting the Rydberg states, phase and intensity noise of the excitation lasers. As these technical errors are reduced, however, the remaining error budget is increasingly set by the finite lifetime of the Rydberg state together with the time required to operate in the blockade regime \cite{Saffman2005Analysis, Saffman2010Quantum, Lib2026Velocity, Tsai2025Benchmarking}. This motivates revisiting the faster interaction-driven gate, which has remained comparatively unexplored because of its sensitivity to the exact interatomic separation. A central question is therefore whether atomic positions can be controlled precisely enough for this gate to reach fidelities at the 99.99~\% level.

The sensitivity to interatomic separation makes the interaction-driven gate a problem of motional control. This is well established in trapped ions, where collective motion mediates entangling gates \cite{CiracZoller1995Quantum, SorensenMlmer1999Quantum, SorensenMolmer2000Entanglement} and can be engineered into non-classical oscillator states \cite{Fluemann2019Encoding, simoni2025nonlinear, Matsos2024Robust}. In neutral atoms, motional control is less mature, but is becoming an increasingly active frontier. Neutral atoms have been cooled near the ground state \cite{Lester2014Raman, Kaufman2012Cooling, lukin_side_band_cooling_thomson}, coherently transported\cite{Beugnon2007Two}, launched and recaptured \cite{Hwang_23_throw_catch}, their motional degree of freedom was squeezed \cite{Lienhard_2025}, and prepared in a motional cat-state \cite{Pampel2026Motional}. Spin-motion coupling has also been used for gate-based cooling \cite{tsai2026gate} and measurement-enabled transfer of spin entanglement to motion \cite{Shaw2025Erasure}. A theoretical Rydberg-gate proposal based on motional geometric-phase acquisition, in analogy with ion-trap geometric-phase gates, has also been put forward \cite{Cozzini2006Fast}.

In parallel, rapid excitation protocols and fast optical-pulse control are pushing neutral-atom platforms toward the ultrafast regime \cite{Chew_2022, Mahesh_2025}. This strongly suggests that achieving sufficiently large and  well controlled Rabi frequencies is now within reach, thereby narrowing the gap between interaction driven and blockade based gates. Moreover, combining such excitation techniques with precise control of atomic motion opens the possibility of realizing interaction driven protocols that are robust against uncertainties in atomic positions. 

This work advances along this direction. We begin in Section~\ref{sec: Overview} by comparing the blockade and interaction-driven gate paradigms, identifying position uncertainty as the inherent limitation of the latter. Motivated by this analysis, we introduce an \textit{ultrafast gate} --- an interaction-driven gate supplemented by a set of coherent motion-control protocols that reduce position-induced errors by nearly two orders of magnitude. First, we show how optical lattices can eliminate uncertainties in the interatomic separation originating from imperfect trap positioning, and describe their implementation and loading in Section~\ref{lattice}. Second, in Section~\ref{motion-echo}, we introduce a \textit{motion-echo} protocol that suppresses the effects of thermal motion within the trapping sites while remaining robust against key experimental imperfections. Third, in Section~\ref{sec:free-flight}, we extend this approach beyond the instantaneous-gate limit to interaction times comparable to the motional dynamics. Finally, in Section~\ref{sec:flux}, we analyze the entangling rates achievable with this architecture and discuss the resulting prospects for ultrafast, high-fidelity Rydberg gates.

\section{Overview of the gate}
\label{sec: Overview}

\subsection{Laser-driven vs interaction-driven gate}

\begin{figure}
    \centering
    \includegraphics[width=\columnwidth]{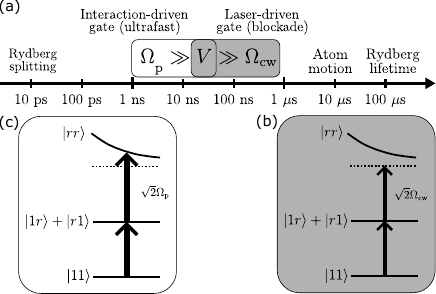}
    \caption{Rydberg gates concepts. (a) Relevant dynamical timescales of the system. Energy-level diagrams in the blockade regime (b) and in the interaction-driven regime (c).
    In (b), the entangling phase is generated by the laser drive, whereas in (c) it is accumulated during the interaction time between the atoms.
    }
    \label{fig1}
\end{figure}

Rydberg-mediated entangling gates can be understood most clearly in two limiting regimes \cite{Jaksch2000Fast}, illustrated in Fig.~\ref{fig1}(a). In both cases, qubit states $\ket{1}$ are coupled to a Rydberg state $\ket{r}$ with Rabi frequency $\Omega$, while the doubly excited state $\ket{\tilde{rr}}$ is shifted by an interaction energy $V$. In the familiar \emph{Rydberg-blockade} regime (Fig.~\ref{fig1}(b)), $V \gg \Omega$, the interaction suppresses simultaneous excitation of both atoms. The gate is then generated by the combination of dynamical and geometric phase accumulated during this laser-driven evolution. Its duration is set by the Rabi timescale and cannot be reduced arbitrarily, since maintaining blockade requires $\Omega \ll V$. In the opposite \emph{interaction-driven} regime (Fig.~\ref{fig1}(c)), $\Omega \gg V$, both atoms are transferred to $\ket{\tilde{rr}}$ and acquire an entangling phase directly from the interaction, $\Phi_{\rm ent}=Vt/\hbar$, reaching the maximal entanglement at $t=\pi\hbar/V$.

\subsection{Fidelity tradeoffs}
For a broad class of error mechanisms, increasing the gate speed is advantageous. The most fundamental is the finite Rydberg lifetime, set by spontaneous emission and black-body-radiation-induced transitions out of the Rydberg manifold \cite{Saffman2010Quantum, Tsai2025Benchmarking, Zhang_2025, Jin_2026}. Since the corresponding error is proportional to the time spent in Rydberg states, faster gates reduce the lifetime contribution. The same general principle applies to several technical and motional errors. Contributions from Doppler shifts, laser phase noise, intensity noise, and slow fluctuations of electric fields all become smaller when the gate is completed in shorter time. In this sense, many of the dominant imperfections favor increasing the rate at which entanglement is generated.

One cannot, however, indefinitely increase the rate of entanglement generation. In blockade gates, the gate time is set by the Rabi frequency $\Omega$, but $\Omega$ cannot be increased independently of the interaction strength $V$. The blockade mechanism requires $V\gg \Omega$, so that the population in the doubly excited state $\ket{\tilde{rr}}$ is strongly suppressed. Driving faster weakens this condition and leads to leakage into the imperfectly blockaded state that scales with powers of $\Omega/V$ \cite{Saffman2010Quantum}. This creates the basic blockade tradeoff --- faster driving reduces errors due to lifetime and slow-noise, but increases blockade error. While pulse shaping, optimal control, and adiabatic protocols can improve robustness to many imperfections \cite{Beterov_2016, Petrosyan_2017, Levine_2019, Jandura_2022, Pagano_2022}, they do not remove the underlying constraint that in the blockade regime, making the gate faster directly weakens the separation of scales on which the blockade approximation relies. Several proposals aim to overcome this limitation using dark states \cite{Petrosyan_2017}, geometric phases \cite{Jin_2024} or even hybrid approaches \cite{Giudici_2025, mostaan_2026}. 

Here, we pursue a complementary route by moving fully into the interaction-driven regime. Rather than suppressing access to the interacting pair state, we deliberately populate it and accumulate the entangling phase directly at rate $V$. For the same interaction strength, the resulting gate can therefore be faster than its blockaded counterpart. The price is a motional tradeoff in place of the blockade one. Because the gate phase derives directly from $V(R)$, the gate inherits the distance sensitivity of the interaction: for a van der Waals interaction, $V \propto R^{-6}$, a position uncertainty $\Delta R$ produces phase noise with infidelity scaling as $\propto (6\Delta R/R)^2$. Faster interaction-driven gates thus reduce lifetime and slow-noise errors at the cost of sensitivity to distance control and motional dynamics. In this work, we show that this limit is less technically challenging than the blockade constraint and can be controlled by engineering the motional response. 

\subsection{The ultrafast gate}
\label{sec: ultrafast gate}

\begin{figure}
    \centering
    \includegraphics[width=\columnwidth]{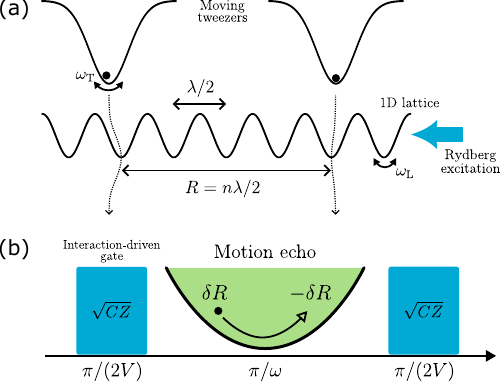}
    \caption{Tools and concepts for interaction-driven gate implementation. (a) Schematic of movable optical tweezers adiabatically loaded into a one-dimensional optical lattice to precisely set interatomic distance $R$. The lattice with period $\lambda/2$ reduces positioning errors when its trapping frequency $\omega_{\rm L}$ exceeds that of the tweezers $\omega_{\rm T}$. (b) Schematic of motion-echo protocol. The CZ gate is implemented in two steps, separated by half a motional oscillation period. In this protocol thermal and quantum fluctuations $\delta R$ of the atomic positions within the trapping potentials cancel to first order.}
    \label{fig2}
\end{figure}

We now turn from the ideal interaction-driven gate to its implementation. We assume shaped laser pulses that maps $\ket{01}$ and $\ket{10}$ to $\ket{0r}$ and $\ket{r0}$, and $\ket{11}$ to a selected interacting pair eigenstate $\ket{\tilde{rr}}$. A controlled-Z (CZ) gate is obtained by accumulating an interaction phase $\phi=Vt/\hbar=\pi$. Under this ideal-transfer assumption, the dominant phase errors arise from the $R^{-6}$ dependence of $V$. We separate them into two contributions: (\textit{i}) static calibration errors $\Delta R$ of the trap separation, and (\textit{ii}) thermal and quantum fluctuations $\delta R$ of the atoms within the traps.

(\textit{i}) \textit{Miscalibration of the distance}:
Considering the gate to be performed in parallel over many pairs of atoms, we require a common interaction time $t$ and inter-atomic distance $R$ over all pairs. A calibration error $\Delta R$ then turns into an error of the entanglement phase $\phi = \pi(1 + \Delta R / R)^{-6}$ propagating to the average gate fidelity as
\begin{equation}
    \mathcal{F} = \frac{7 - 3\cos{\phi}}{10} \sim 1 - \frac{3}{20} \left(6 \pi  \frac{\Delta R}{R}\right)^2,
    \label{eq:fidelity_phase_uncertainty}
\end{equation}
where the prefactor captures the fact that only $\ket{11}$, but not $\{\ket{00},\ket{10},\ket{01}\}$, experiences the error (derived in Appendix ~\ref{app: derivationSystPhaseError}).  

Targeting a gate error $1 - \mathcal{F} \lesssim 10^{-4}$, at a typical interatomic distance $R \sim 3~\mu\text{m}$, translates into asking for $\Delta R < 4~\text{nm}$. To highlight the difficulty of this requirement, we can compare it to the typical size (waist) of the optical tweezers of $500 - 1000$~nm. 
We could consider to measure precisely the distance between optical traps and adjust them accordingly, as already demonstrated for holographic tweezers~\cite{Chew_2024}. However, our experience is that positioning at such precise level would be technically challenging and require frequent (minutes to hours) re-calibration.

(\textit{ii}) \textit{Position fluctuations}:
Even if the traps could be perfectly positioned, there is still the important issue of the atoms moving inside the traps. These fluctuations $\delta R$ can originate from an excess thermal energy --- due to imperfect laser cooling of the atoms after loading in the traps or from heating by previous operations (state preparation, moves, gates) --- and ultimately from quantum fluctuations. The gate error from such fluctuations can be similarly estimated from Eq.~(\ref{eq:fidelity_phase_uncertainty}), a strict derivation is detailed later, putting the same stringent constraint $\delta R < 4$~nm. 

For comparison, the quantum uncertainty of distance between two atoms (root-mean-square spatial extent of the relative motional ground state) is given by $\delta R_0=\sqrt{\hbar/(2\mu\omega)}$, with reduced mass of the two-atom system $\mu=m/2$ and trapping frequency $\omega$.  For the mass of rubidium, $m$, and a typical trap frequency of $\omega_T = 2\pi \times 100$~kHz for optical tweezers, we have $\delta R_0 = 34$~nm giving $\mathcal{F} \simeq 99.3\%$, below the performance of current state-of-the-art blockade gate. 
In addition, we aim for gate performance that does not require atoms to be perfectly cooled to the motional ground state, but instead remains compatible with a mean phonon number up to $\bar n \approx 3$.
The fluctuation could be reduced by simply increasing laser power in the optical tweezers, but the scaling law is not favorable ($\delta R \propto P^{-1/4}$). A $5000\times$ deeper tweezers would be needed to just have the quantum fluctuation at the requested level, and an additional factor of 3-10 to make the gate robust to additional thermal fluctuation of the atoms. Without increasing trap depth, one can consider squeezing the position uncertainty beyond the standard quantum limit, as demonstrated in Ref~\cite{Lienhard_2025}, but the tweezers anharmonicity prevented reaching the above requirement. 

These considerations regarding error channels (\textit{i}) and (\textit{ii}) call for new mitigation strategies to enable the interaction-driven gate.
To suppress (\textit{i}), we propose to add a two-dimensional optical lattice on top of the array of tweezers, as illustrated in Fig.\ref{fig2}(a). The lattice, set to be stronger than the tweezers, forces the atoms to localize at the exactly periodic minima of the lattice potential, providing a stable and accurate calibration of the interatomic distance. 
The method is highly parallel and scalable due to the long axial extent of the lattice, and directly compatible with the zone-architecture of neutral-atom quantum computer. 
We also note that this technique facilitates calibration of the dynamical entangling phase compared to the blockade gate, whose geometric entangling phase require a precise intensity of the Rydberg excitation laser, challenging to maintain homogeneous over a large area.

To suppress (\textit{ii}), we introduce a \textit{motion-echo} protocol, depicted in Fig.\ref{fig2}(b). The CZ gate is divided into two successive operations (each corresponding to a $\sqrt{\rm CZ}$), separated by half a period of the motion of atoms in the harmonic trapping potential defined by the combined tweezers and lattice. In essence, a distance error $\delta R$ occurring during the first $\sqrt{\rm CZ}$ operation is reversed to $-\delta R$ after evolving for half a period, so that the coherent errors accumulated in the two halves cancel through destructive interference. We will show that this coherent control technique reduces gate errors to a sub-$10^{-4}$ level, even when considering thermal excitation of the atoms in the traps up to $\bar{n} \simeq 3$, \textit{i.e.}, we do not require the atoms to be in the motional ground-state to achieve this level of fidelity. Although the complete protocol has a longer duration, the cumulative time spent in the Rydberg state remains short, thereby limiting errors associated with the finite Rydberg-state lifetime.

\section{Lattice for calibrated gate}
\label{lattice}

To mitigate uncertainties in distance calibration, we propose using the spatial periodicity of an optical lattice. We first introduce the basic mechanism using a one-dimensional lattice along the interaction axis $x$, which pins the atoms to sites separated by integer multiples of $\lambda/2$ and suppresses residual tweezer-position errors along $x$. We then show that, once this longitudinal calibration error is reduced, the dominant remaining contribution arises from transverse position fluctuations along the weakly confined $z$-axis of the tweezers. This motivates extending the scheme to a two-dimensional lattice, which preserves the calibrated spacing along $x$ while adding transverse confinement along $z$. 

\subsection{One-dimensional lattice: longitudinal distance calibration}

\begin{figure}
    \centering
    \includegraphics[width=\columnwidth]{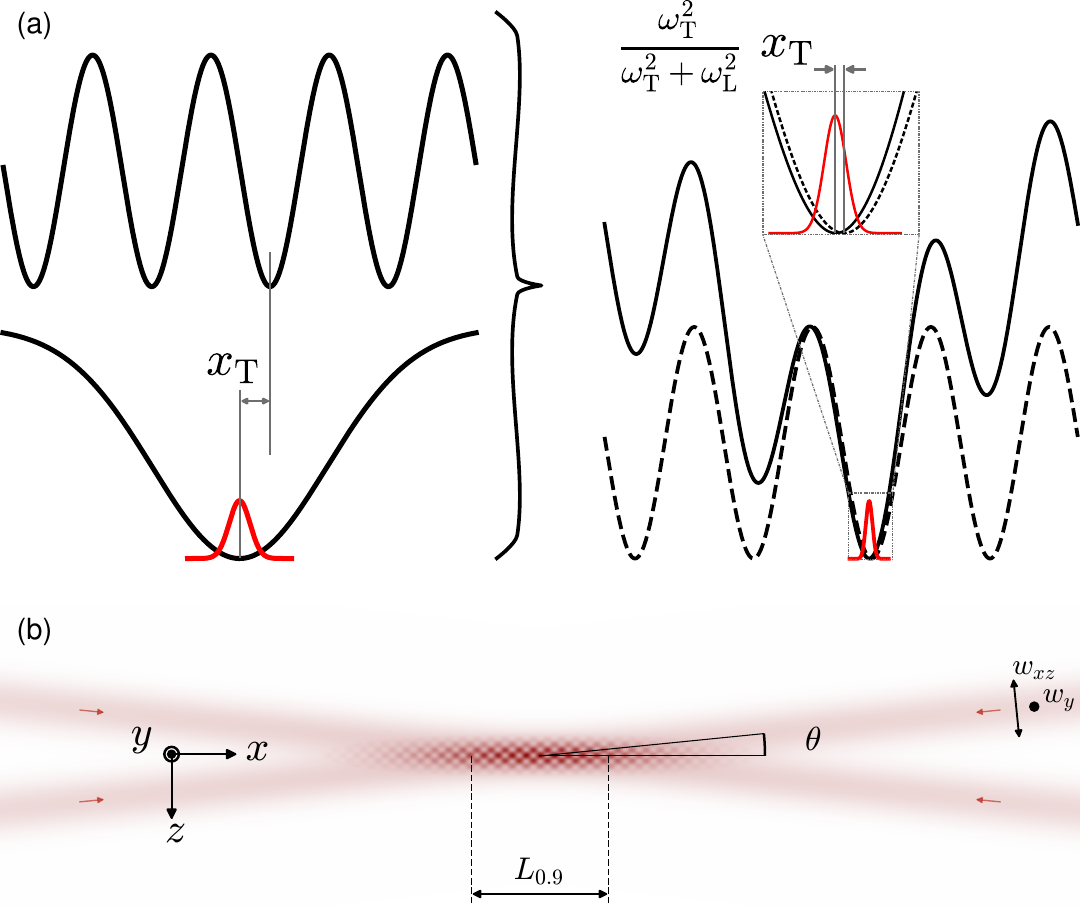}
    \caption{
    (a) Combined tweezer–lattice trapping potential for a tweezer displaced by $x_T$ from the lattice site. The resulting total potential shows that the shift of the trap minimum is strongly suppressed when $\omega_{L}\gg\omega_{T}$. Ground-state harmonic oscillator wavefunctions are shown in light red, illustrating the reduction of quantum fluctuations $\delta R_0$ at the lattice site. (b) Geometry of the two-dimensional lattice. Two standing waves are generated by elliptical beams propagating in the $x$-$z$ plane at angles $\pm\theta$ with respect to the interaction axis $x$. The resulting summed potential confines the atoms along both $x$ and $z$. Near the central axis the trap-frequency ratio is $\omega_{z, L}/\omega_{x,L}=\tan\theta$. The in-plane elliptical waist $w_{xz}$ determines the usable longitudinal extent, with $L_{0.9}$ indicating the region over which the lattice frequencies remain within $10\%$ of their peak values.
    }
    \label{fig:Lattice plus Tweezer potential}
\end{figure}

In the one-dimensional implementation, two counter-propagating lattice beams of wavelength $\lambda$ generate a standing wave with spatial period $\lambda/2$. The distance between two atoms can then be set extremely precisely to $R = n \times \lambda / 2$ ($n$, an arbitrary integer) by superposing both the tweezers and lattice potential as shown in Fig.~\ref{fig2}(a). Throughout this paper, we fix this distance to $R = 3~\mu\text{m}$.
Note that the tweezers cannot be switched off, as it provides the confinement along $y$ and $z$, the one-dimensional lattice acting only along the $x$-direction. As such, a control error $x_T$ in the position of the tweezers with respect to the lattice site can still perturb the positioning of the atoms, as shown in Fig.~\ref{fig:Lattice plus Tweezer potential}(a). By choosing the lattice trap frequency to be much larger than the tweezer frequency $\omega_{L}\gg\omega_{T}$, the lattice dominates and effectively pins the atoms to the correct position. 

More formally, along the direction $x$, the combined potential can be written as:
\begin{align}
    \hat{U} & \simeq \frac{1}{2}m \omega_T^2  \left( \hat{x} - x_{T} \right)^2 + \frac{1}{2}m \omega_{L}^2 \hat{x}^2 \nonumber\\
    & \simeq \frac{1}{2} m \left(\omega_{T}^2 + \omega_{L}^2 \right) \left( \hat{x} - \frac{\omega_{T}^2}{\omega_{T}^2 + \omega_{L}^2} x_{T}\right)^2
\end{align}
where, the first term is the misplaced tweezers potential, and the second term is the lattice confinement. For both, we keep only the harmonic, quadratic, approximation. In the second line, we see that the error in position of the local minimum of the combined potential is suppressed quadratically with the ratio of trapping frequencies, such that by choosing $\omega_L = 10 \times \omega_T \simeq 2 \pi \times 1$~MHz, we obtain a reduction of the control error by a factor of 100. Requiring the tweezers position to be within $\vert x_T \vert < \lambda/4 \sim 200$~nm of the lattice minima, which is needed to load the atom in the correct lattice site, we extract an upper bound of $\pm 2$~nm for the position error of each trap. The combined tweezers+lattice thus provides a worst-case inter-atomic distance error $\Delta R = 4$~nm, at the threshold required to bring the gate error due to miscalibration of the tweezers along $x$ below $10^{-4}$. 

Achieving a lattice trapping frequency of $\omega_L = 2\pi \times 1$~MHz is experimentally accessible. For example, in the case of $^{87}$Rb, this trap frequency is obtained with two counter-propagating beams at $\lambda = 775~\text{nm}$ (blue-detuned from the D2 transition by 2.5~THz), each with a power of $30~\text{mW}$ and a waist of $w_0=10~\mu\text{m}$.
The extended axial ($x$-direction) profile of this lattice also supports several atom pairs at the same time, enabling parallel two-qubit gate operations. More specifically, the limit is set by the decrease of the lattice trapping frequency $\omega_L \propto 1/\sqrt{1 + (x/z_R)^2}$ as we move away from the lattice waist location, due to the finite Rayleigh length $z_R = \pi w_0^2/\lambda \sim 400$~$\mu$m. Restricting the atoms to $|x|\lesssim0.5z_R$ reduces $\omega_L$ by only about $10\%$, while preserving the hierarchy $\omega_L\gg\omega_T$. As discussed in Sec.\ref{sec:timing}, this range is also sufficient for the more demanding motion-echo protocol. The corresponding axial region, of order $400~\mu\mathrm{m}$, already spans the typical entangling zone of current neutral-atom processors, and the favorable scaling $z_R \propto w_0^2$ enables to easily increase it for future larger systems. 

We also note that the tighter lattice confinement also reduces the extent of the ground-state wavefunction $\propto \omega^{-1/2}$ (see Fig.~\ref{fig:Lattice plus Tweezer potential}(a)), thereby decreasing the error from thermal and quantum fluctuations. With our parameters, we now have $\delta R_0 = 11$~nm and an average gate fidelity of $99.93$\,\% for an atom in the motional ground-state, getting closer to the $99.99\,\%$ target. In the Sec.~\ref{motion-echo}, we present a second additional protocol to reach this goal and, importantly, make the gate robust to thermal excitation. 

\subsection{Two-dimensional lattice: transverse positioning and thermal fluctuation errors}
\label{sec:transverse_errors}

While the one-dimensional optical lattice strongly pins the atomic positions along the inter-atomic axis ($x$) it provides no additional confinement along the transverse axes ($y$ and $z$). Consequently we consider the impact of transverse positioning errors, and error arising from thermal and quantum fluctuations.

Transverse positioning of two atoms perturbs the inter-atomic distance to second order as $\Delta R \simeq (\Delta y)^2/2R$. To maintain the interaction-phase error below our threshold, the rms transverse position error of the tweezers must be kept below 110~nm. This is easily achievable along the tightly focused $y$-axis, but is less trivial along the $z$-axis where the tweezers Rayleigh length is of a few micrometers. We have recently demonstrated holographic mitigation technique to measure and adjust the $z$-position of static tweezers from an rms error of 150~nm to 75~nm~\cite{Chew_2024}, thus also clearing this requirement. Although the proposed technique applies to static tweezers, the dynamically moving tweezers required for the transport of the atoms into the lattice, defined by an acousto-optic deflector, are expected to exhibit lower $z$ fluctuations (we suspect that it is the hologram imperfection that gave rise to most of the $z$ error in the first place). 

In addition to static misalignments, finite transverse confinement introduces quantum and thermal positional fluctuations. Because the transverse trapping frequencies are set only by the tweezers and are typically $\omega_{y (z), T} \simeq 2\pi \times 100 \, (25)$~kHz, for a diffraction-limited Gaussian tweezers ($\lambda = 800$~nm, $w = 0.7 \, \mu$m), the position fluctuations are significantly larger than along $x$-axis. Furthermore the echo sequence (that will be discussed in Sec.~\ref{motion-echo}) is ineffective in canceling the transverse position fluctuations as it is timed for the much larger lattice trapping frequency. 

Focusing on the weakest confinement along the $z$-direction, the second-order perturbation to the interaction distance, derived in Appendix \ref{app: Transverse fluctu}, yields an average gate error for a thermal state given by 
\begin{equation}
    \label{eq:transverse_positioning_error_fluctuation}
    1-\mathcal{F} \simeq \frac{9}{20} \left( 3\pi \frac{\delta Z_0^2}{R^2} \right)^2(2 \bar n_z + 1)^2,
\end{equation}
where  $\delta Z_0$ is the ground state spatial extent of the relative $z$ position, and $\bar n_z$ is the mean occupation number.
For the weak tweezer axis trap frequency $\omega_{z,T}$, we get $\delta Z_0 \simeq 68$~nm. This leads to a negligible gate error of $10^{-5}$ for atoms perfectly cooled to the motional axial ground state and of $10^{-4}$ for $\bar n_z \simeq 1.3$. This imposes strict cooling requirement along the weakly confining tweezer axis. To maintain high fidelity while accommodating a more realistic thermal occupation of $\bar n_z \leq 3$, we propose to add confinement along $z$ to bring the trapping frequency to $\sim 100$~kHz. This can be done for instance by adding an optical lattice in that plane \cite{Norcia_2024}. 

A weak transverse lattice can be incorporated without adding a separate beam axis by replacing the single standing wave along $x$ with two standing waves in the $x$-$z$ plane, tilted by angles $\pm\theta$ with respect to the $x$-axis. Near the central axis, the sum of the two standing-wave potentials provides confinement in both directions, shown in Fig.~\ref{fig:Lattice plus Tweezer potential}(b). The corresponding ratio of lattice trap frequencies is $\omega_{z,L}/\omega_{x,L}=\tan\theta$. Thus, targeting $\omega_{x, L}=2\pi\times1~{\rm MHz}$ and $\omega_{z,L}=2\pi\times100~{\rm kHz}$ requires only $\theta=\arctan(0.1)\simeq5.7^\circ$. This transverse confinement comes at the cost of a reduced longitudinal region over which the lattice intensity is sufficiently uniform. For circular beams with waist $w_0=10~\mu{\rm m}$, the region in which both lattice trap frequencies remain within $10\%$ of their central values is $L_{0.9}\simeq6.5w_0\simeq65~\mu{\rm m}$, as derived in Appendix~\ref{app:2d_lattice}. This is shorter than the $\sim400~\mu{\rm m}$ axial region available for the one-dimensional lattice considered above. The uniform region can be extended by using elliptical beams with a larger waist $w_{xz}$ in the $x$-$z$ plane. In this configuration recovering a $L_{0.9}=400~\mu{\rm m}$ entangling zone extent requires $w_{xz}\simeq88~\mu{\rm m}$, as derived in Appendix~\ref{app:2d_lattice}. For the same detuning and target longitudinal trap frequency, this corresponds to a power of approximately $140~{\rm mW}$ per beam. Note that, compared to the estimate in the previous section, the longitudinal confinement is now provided by two tilted standing waves instead of one, so each standing wave requires only half the depth.

\subsection{Near Adiabatic transport into the lattice}
\label{near_adiabatic_transport}

\begin{figure}
    \centering
    \includegraphics[width=1\linewidth]{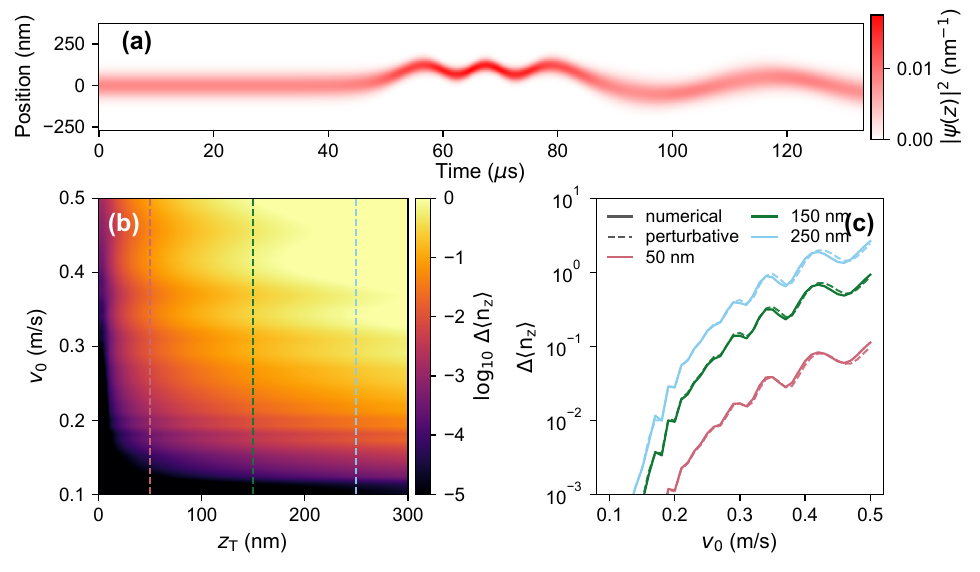}
    \caption{Transport-induced motional excitation in the lattice gate zone. (a) Probability density of an atom transported into and out of the lattice interaction region at $v_0=0.6~\mathrm{m/s}$ with tweezer–lattice offset $z_T=100~\mathrm{nm}$. The atom is pinned to the lattice minimum while traversing the interaction region; the residual oscillation after the passage reflects motional excitation. (b) Change in the mean vibrational occupation $\Delta\langle n_z\rangle$ after the full transport sequence, obtained from numerical propagation in the full time-dependent potential and averaged over a $10\%$ Gaussian spread in lattice trap frequency. Atom is initially assumed to be in the motional ground state. (c) Numerical cuts from panel (b) compared with the perturbative prediction of Eq.~\eqref{eq:v_approx}. The harmonic theory accurately describes the small-offset regime.}
    \label{fig: adiabatic transfer}
\end{figure}

Transport into and out of the lattice is realized by near-adiabatically dragging the atoms into the two-dimensional lattice with a tweezer moving at constant speed $v_0$ along the transverse, $y$-direction. In general, the tweezer minimum is displaced from the target lattice site by an offset $(x_T, z_T)$, due to the imperfect calibration of the two potential minima. Fig.~\ref{fig: adiabatic transfer} (a) shows a representative evolution of atomic probability density for $v_0=0.6~\mathrm{m/s}$, $x_T=0~\mathrm{nm}$, and $z_T=100~\mathrm{nm}$. As the tweezer crosses the lattice region, the stronger lattice confinement pins the atom to the lattice minimum. For finite offset $z_T$, however, sufficiently rapid transport induces motional excitation, dominated by a residual displacement of the atomic wave packet which appears as a residual oscillatory motion of the probability density after the passage.

Because the trap frequency of the tweezer is significantly stronger along $x$ than along $z$, the dominant transport-induced excitation arises from the weaker $z$ direction. We therefore focus below on the one-dimensional dynamics along $z$, which sets the limiting scale for adiabatic transfer. The analysis of the dynamics along $x$ is analogous and gives a weaker constraint. To quantify the effect of dragging an atom through the lattice, we numerically propagate the motional wave function by directly solving the Schr\"odinger equation in the full time-dependent one-dimensional potential along each of the lattice axis (for more details see Appendix~\ref{app:anharmonic_extensions}). For small offsets $z_T$, the wave packet samples only a limited region of the tweezer potential, such that the combined confinement can be approximated as harmonic. In this regime, the dynamics admit a simple analytical description in place of the full physical potential. Along the $z$-axis the corresponding Hamiltonian is $\hat H_z(t)=\hat p_z^{2}/(2m)+\frac{1}{2}m\omega_z^{2}(t)\left[\hat z-z_{0}(t)\right]^{2}$,
where $\omega_z^{2}(t)=\omega_{z,T}^{2}+\xi(t)\omega_{z,L}^{2}$ and
$z_{0}(t)=\omega_{z,T}^{2}z_{T}/\omega_z^{2}(t)$.
In the perturbative regime, the change in the mean vibrational occupation is
$\Delta\langle n_z\rangle \simeq |\nu_z|^{2}+|\beta_z|^{2}$, with
\begin{align}
\nu_z &\simeq e^{-i\phi_z(T)} \int_{0}^{T}dt\frac{\dot\omega_z(t)}{2\omega_z(t)}e^{2i\phi_z(t)},\nonumber\\
\beta_z &\simeq -e^{-i\phi_z(T)}\int_{0}^{T}dt\frac{\dot z_{0}(t)}{2\delta_z(t)}e^{i\phi_z(t)},
\label{eq:v_approx}
\end{align}
with $\nu_z$ and $\beta_z$ describing the squeezing and displacement contributions along the $z$-axis, respectively. Here $\phi_z(t)=\int_{0}^{t}\omega_z(t')dt'$ and $\delta_z(t)=\sqrt{\hbar/\left[2m\omega_z(t)\right]}$.
The derivation is presented in Appendix~\ref{app:diabatic_transport}. 

The numerical simulations reveal narrow parameter regions in which excitation is strongly suppressed even at relatively high transport speeds (see Eq.~\eqref{eq:v_approx}). These minima, however, do not provide robust operating points, as their location is sensitive to both $\omega_{z,L}$ and $\omega_{z,T}$, while parallel operation across many sites inevitably samples a distribution of lattice trap frequencies. We therefore characterize the transport by averaging the final excitation over a Gaussian distribution of lattice frequencies with a $10\%$ standard deviation. Figure~\ref{fig: adiabatic transfer}(b) shows the resulting numeric change in the mean vibrational occupation of the atom in the tweezer after a complete in-and-out passage. As expected, the excitation grows with both the transport velocity and the tweezer–lattice offset. Figure~\ref{fig: adiabatic transfer}(c) compares selected numerical cuts from Fig~\ref{fig: adiabatic transfer}(b) with analytical predictions of Eq.~\eqref{eq:v_approx}. For the relevant range of offsets $z_T$, the analytical treatment quantitatively reproduces the numerical results. Along the $x$-direction, the perturbative approximation begins to deviate from the full numerical calculation for offsets larger than $100~\mathrm{nm}$, where anharmonicity of the tweezer becomes important. Requiring a mean excitation of at most $\bar n_z \sim 10^{-2}$ per passage yields an allowable transport speed of approximately $v_{\text{max}} \sim 0.2~\mathrm{m/s}$ for a maximal offset of $z_T \sim 150~\mathrm{nm}$.

\section{Motion-echo for position uncertainty}
\label{motion-echo}

We now treat the errors coming from the motion of the atoms within the traps. Since we consider a two-atom system, we separate the center-of-mass and relative motion, and only focus on the relative degree-of-freedom which couples to the van der Waals potential. We write the interatomic distance operator as $\hat{R}=R+\delta\hat{R}$ and perform a Taylor expansion of the van der Waals potential around the average distance $R$ giving:
\begin{equation}
    \hat{H}_\text{int} =  V \left[ 1 - 6 \frac{\delta \hat{R}}{R} +  21 \left( \frac{\delta \hat{R}}{R}\right)^2 + ... \right], 
    \label{eq:H_Int}
\end{equation}
where we keep terms up to second order, the higher ones being not relevant for our experimental parameters. 
The zeroth-order term is the energy shift $V$ used to realize the interaction-driven gate in a time $t = \pi\hbar/V$. The first-order term is the gradient of potential around the mean interatomic distance, \textit{i.e.}, the van der Waals force $F = -\text{d} V_{vdW}/\text{d}R$.  
The second-order term is the curvature of the potential \cite{Chew_2022, Bharti_2024, Nill_2025}. These position-dependent terms acts only on the state $\ket{11}$, and not on the others, giving rise to spin-motion entanglement, and then, when tracing over the motional degree-of-freedom, a reduced gate fidelity. 

For an initially pure motional state $\ket{\psi_m}$, the average gate fidelity \cite{Nielsen_2002} can be expressed as
\begin{equation}
    \mathcal{F} = \frac{7 + 3\Re(s)}{10}
    \label{eq:fidelity spin motion}
\end{equation}
where $s = \bra{\psi_m} \exp{(-i \hat{H}_\text{int} t /\hbar )}\ket{\psi_m}$ denotes the wavefunction overlap between the motional state in the spin-sector unaffected by the gate ($\{\ket{00},\ket{10},\ket{01}\}$) and  $\ket{11}$ which experiences the position-dependent van der Waals interaction $\hat{H}_\text{int}$. The formula is generalized in Appendix~\ref{app: FidelyDer} to arbitrary mixed states.

\begin{figure*}
    \centering
    \includegraphics[width=\textwidth]{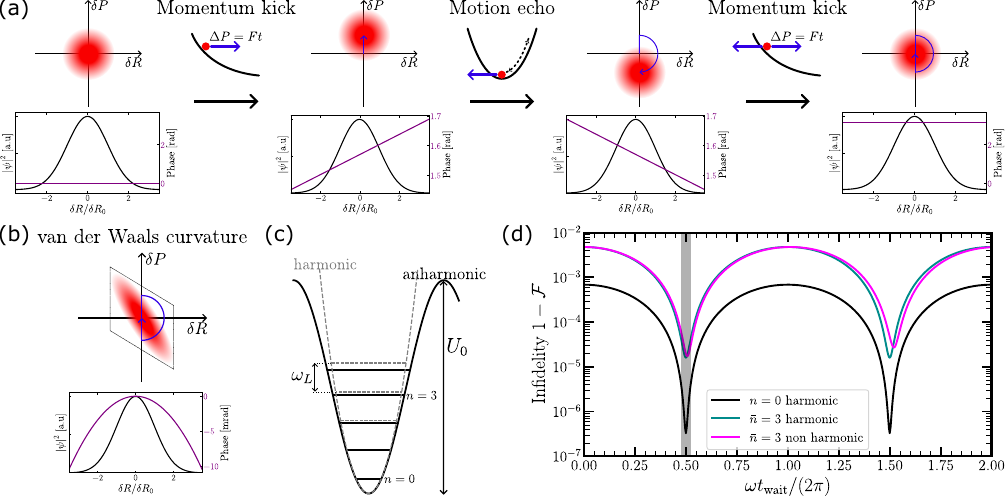}
    \caption{Motion echo sequence. (a) Schematic of the interaction gate. Each panel shows both a phase-space representation and the corresponding position-space probability density, with the associated phase profile. The initial state has zero phase. After the first interaction step, a global $\pi/2$ phase is acquired and the state is displaced along $\delta P$ axis, corresponding to a linear phase gradient. The third step shows the state after a half period rotation in the harmonic trap. The final interaction adds an identical phase, reaching a total of $\pi$, and cancels the phase gradient, returning the state to its initial point in phase space. (b) Same  sequence including the second order term in the Taylor expansion of van der Waals potential. The final state is sheared along the $\delta P$ axis, corresponding to a residual quadratic phase gradient. (c) Schematic of the lattice trapping potential, illustrating its anharmonic nature, along with its harmonic approximation (dotted gray line). The corresponding energy levels are shown; note that the lattice level spacing is non-uniform. (d) Gate infidelity as a function of the rotation angle in the trap between two interactions steps, shown for the motional ground state in an harmonic trap (black), a thermal state $\bar n =3$ in a harmonic trap (turquoise) and a nonharmonic trap (pink). The gray overlay indicates the range of angles corresponding to variations of the trapping frequency over $L_{0.9}$}
    \label{fig:MotionEchoScheme}
\end{figure*}

It is helpful to visualize the wavefunction, along the $x$-direction defined by the lattice, in both real-space and phase-space, as shown in Fig.~\ref{fig:MotionEchoScheme}(a) for the atoms initially in the motional ground state. 
When experiencing $\hat{H}_\text{int}$, the state $\ket{11}$ accumulates the entangling phase from the van der Waals shift $V$, but also experiences the van der Waals force $F$ imparting a momentum kick $\delta P = F t = 3 h/R \simeq \mu \times (9.2 ~\text{mm/s})$ (as observed in \cite{Emperauger_2025}), where $t = \pi \hbar/V$ is the interaction time required to implement a CZ-gate. 
The momentum kick appears as a linear phase gradient over the wavepacket in real-space, and as a displacement in phase-space. Note that, in  Fig.~\ref{fig:MotionEchoScheme}(a), the displacement is highly exaggerated for visualization, as its amplitude is only $\sim 14 \, \%$ of the ground state momentum uncertainty $\delta P_0 = \sqrt{\hbar \mu \omega_{x,L}/2} \simeq \mu \times$~(67 mm/s). 
The overlap of the displaced state with the initial ground state reads $s \sim \exp(-(6\pi \delta R_0/R )^2/2)$, which, when expanded to first-order and injected in Eq.~(\ref{eq:fidelity spin motion}), gives the same expression for the fidelity as in Eq.~(\ref{eq:fidelity_phase_uncertainty}). 
Note that we have performed the ``frozen'' approximation where we assumed that the atoms are not moving during the interaction time $t$. This corresponds to the limit $V/\hbar \gg \omega_{x,L}$: the entangling time is much shorter than the period of oscillation in the trap. We will go beyond this approximation in Section~\ref{sec:free-flight}. 

To mitigate the errors arising from position fluctuations we propose the motion-echo protocol, sketched in Fig.~\ref{fig:MotionEchoScheme}(a). The CZ gate is split in two consecutive $\sqrt{\text{CZ}}$-gates each of duration $t = \pi\hbar/(2V)$ separated by a delay of half a trapping period during which the atoms oscillate in the lattice (note that the atoms are in the electronic ground states, and not in the Rydberg orbits, during this time). Following the initial momentum kick (second panel), evolution in the harmonic trap flips the displacement and the associated phase gradient (third panel). The second interaction pulse then applies an equal and opposite momentum kick, canceling the net momentum transfer and restoring the state to its initial point in phase space with a uniform flat phase profile across the wavepacket (final panel). 

This protocol corrects the first-order sensitivity of the interaction gate to the position error along $x$. It works for an atom in an arbitrary initial state, and is not limited to the motional ground state. In the following sections, we will check its robustness to timing error of the echo, curvature of the van der Waals potential and anharmonicity of the trapping potential. 

\subsection{Timing error of the echo}
\label{sec:timing}

The motion-echo requires to apply the second interaction pulse after exactly half a period of oscillation. From now on, we consider solely the dynamics along the $x$-axis within the lattice and therefore drop the index on the trapping frequency, writing $\omega \equiv \omega_{x,L}$. Since pairs of atoms experience different trapping frequencies $\omega$ along the lattice, due to its finite Rayleigh length discussed in Section~\ref{lattice}, there will be slight variation of timing $\theta = \omega t_{\rm wait} \neq \pi$ giving rise to a gate error. 

We introduce the momentum kick, in reduced unit, $\kappa\equiv 6 \phi~\delta R_0/R$, where $\phi\equiv Vt/\hbar$ is the entangling phase ($\pi/2$ for the $\sqrt{\mathrm{CZ}}$ operation). 
The total propagator of the echo sequence acting on the relative motional Hilbert space then reads 
\begin{align}
\hat U &= e^{-\frac{i}{\hbar}\hat H_\text{int}t} \hat R(\theta) e^{-\frac{i}{\hbar}\hat H_\text{int}t}\\ \notag
&=e^{-i2\phi}e^{i\Phi_{\rm geo}}
\hat D \left(\alpha_{\rm res}\right)\hat R(\theta),
\label{eq:Uecho_general}
\end{align}
where the first term realizes the CZ gate, the second term  $\Phi_{\rm geo}=\kappa^2\sin\theta$ is a geometric phase, the third term $\hat D \left(\alpha_{\rm res}\right)$ is a displacement operator with complex amplitude $\alpha_{\rm res}=i\kappa\left(1+e^{-i\theta}\right)$, and the last operator $\hat R(\theta)$ is a phase-space rotation by $\theta$. 

For a perfect echo ($\theta=\pi$), the displacement and the geometric phase cancels. For a small error $\Delta \theta$, there is a residual displacement $\alpha_{\rm res} \simeq - \kappa \Delta \theta$ (along the position-axis), and geometric phase $\Phi_{\rm geo}\simeq -\kappa^2\Delta\theta$, which decreases the fidelity for an atom in the motional ground state as
\begin{equation}
1- F \simeq \frac{3}{20} \left(|\alpha_{\rm res}|^2
+|\Phi_{\rm geo}|^2 \right)
\simeq \frac{3}{20}(\kappa\Delta\theta)^2,
\end{equation}
where, in the last approximation, we only kept the leading order term in $\kappa \simeq 0.035 \ll 1$. 
We show this infidelity as a function of the echo timing as the black curve in Fig.~\ref{fig:MotionEchoScheme}(d). Without echo ($\theta = 0$), the error is $7 \times 10^{-4}$, as calculated previously. For a perfect motion-echo ($\theta = \pi$, or odd multiples of $\pi$), the error is strongly suppressed, and does not vanishes completely only because of the curvature of the van der Waals potential that we discuss later in Section~\ref{sec:curvature}. Away from the ideal echo, the error increases quadratically with $(\Delta \theta) ^2$, and to keep it below $10^{-4}$, the trapping frequency must be known to better than $\sim 20~\%$. We recall that for atoms located within $\pm 0.5 z_R$ of the lattice waist, the lattice trapping frequency decreases by only 10~\% from the center to the edge. By calibrating the echo for the median trapping frequency, the variation of timing is only $\pm 5$~\%, which we show as the grayed-out region in Fig.~\ref{fig:MotionEchoScheme}(d). 

While the timing error is not an issue at the $10^{-4}$ level for atoms in the motional ground state, it becomes more relevant at finite temperature. Introducing the mean motional occupation $\bar{n}$, the infidelity now reads (see Appendix \ref{app: DerivationEchoSequence}):
\begin{equation}
    \label{eq:residual_echo_error}
    1-\mathcal{F} \simeq \frac{3}{20}\kappa^2\Delta\theta^2 \left(2\bar n + 1 \right).
\end{equation}
Consequently, the tolerance to timing error $\Delta \theta$ decreases as $(2\bar n+1) ^{-1/2}$. This trend is illustrated, for a representative $\bar n = 3$, by the turquoise curve in Fig.~\ref{fig:MotionEchoScheme}(d). The error remains below $10^{-4}$ over the grayed-out region, but there is almost no more margin, which thus sets our requirement for atoms to be colder than $\bar n \leq 3$. For reference, (Raman-)sideband cooling can routinely prepare $\bar n < 0.1$, and even the faster EIT cooling reaches $\bar n \sim 1$ \cite{Chow_2024, Chiu_2025, Bluvstein_2026}. The limit $\bar n \leq 3$ is thus well within the state-preparation capabilities of current platform, and will only limit the number of operation that can be performed before re-cooling needs to be applied. 

\subsection{Curvature of the van der Waals potential}
\label{sec:curvature}

We now turn to the next order of position-dependence of the van der Waals potential (remember Eq.~(\ref{eq:H_Int})). While the motion-echo is successful in canceling the first-order term, the van der Waals force; it does not work on the second-order term: the curvature of the van der Waals potential: $21 V \delta \hat R^2/ R^2$. 

This term captures that the force ($\propto R^{-7}$) acting on the wavefunction is position-dependent: at shorter inter-atomic distance, the force is stronger and gives a larger momentum kick. Quantitatively, over the typical size of the motional ground-state $\delta R_0$, the force varies by $7 \delta R_0 / R \simeq 2 \, \%$.  This leads to a shearing of the state in phase space, illustrated (not to scale) in Fig.~\ref{fig:MotionEchoScheme}(b), and leaves the relative motional wavepacket in a slightly squeezed state~\cite{Bharti_2024}. The squeezing cannot be removed by the motion-echo: after half a rotation in the trap, the orientation of the squeezed state is the same, and the second $\sqrt{{\rm CZ}}$-pulse thus contribute the same squeezing as the first interaction pulse. It can also be seen in real-space, the echo flips $\delta \hat{R} \rightarrow - \delta \hat{R}$, which does nothing on this quadratic term  $(\delta \hat{R})^2$. 

The curvature of the van der Waals potential can thus limit the performance of the interaction-driven gate. 
The expression for the gate error due to the curvature is: 
\begin{equation}
    \label{eq:residual_van_der_valse}
    1-\mathcal{F} \simeq \frac{9}{20} \left(21 \pi  \frac{\delta R_0^2}{R^2}\right)^2 \left(2 \bar n + 1 \right)^2.
\end{equation}
For an atom in $n = 0$, the contribution is a negligible $3\times 10^{-7}$, as can be read on the black curve in Fig.~\ref{fig:MotionEchoScheme}(d), when the motion-echo removes the dominant error from the first-order term.

As the error increases quadratically with $n$, it becomes more relevant for higher Fock states. It goes above our target for $\bar{n} \simeq 8$ for a perfect motion-echo, and $\bar{n} \simeq 5$ if we include the  $\pm 5~\%$ motion-echo timing error. At our specified $\bar n \leq 3$, its contribution remains at $5 \times 10^{-5}$ (Fig~\ref{fig:MotionEchoScheme}(d), turquoise curve), allowing us to state that it does not limit our proposal. We conclude by pointing out that the next-order term $\propto (\delta \hat{R})^{3}$ is odd and thus canceled by the motion-echo, and that all higher-order even terms are further negligible with respect to the curvature that we treated here. 

\subsection{Anharmonicity of the trap}
\label{sec:anharmonicity}

The motion-echo protocol supposes that the trapping potential is harmonic, such that it flips $\delta \hat{R} \rightarrow -\delta\hat{R}$ at $\omega t_{\rm wait} = \pi$, irrespective of the exact motional state. 
However, the lattice potential is a sinusoid, for which the harmonic approximation will fail for higher excited motional states, as shown in Fig.~\ref{fig:MotionEchoScheme}(c). 

This is captured quantitatively by expanding the lattice potential to fourth order in $\hat{x}$: 
\begin{equation}
    \hat U = \frac{1}{2} m \omega^2 \hat x^2 - \frac{m \omega^2 k^2}{6} \hat x^4
\end{equation}
with $k = 2\pi/\lambda$. The quartic perturbation decreases the energy levels compared to the one of an ideal harmonic potential as $\hbar^2 k^2/8m \times (2n^2 +2n+1)$, such that, as an example, the state $n =3$ of the lattice is depressed by $24$~kHz for our parameters. As this remains small compared to the trapping frequency, it validates that we restrict the expansion of the lattice potential to fourth order for the low motional state considered here.  

We included this perturbation in numerical simulations of the motion-echo to estimate its impact on our proposed protocol, as shown in the pink curve of Fig~\ref{fig:MotionEchoScheme}(d). First, the optimal echo occurs at slightly longer times compared to the purely harmonic case, as expected intuitively from the slower dynamics for higher excited states. Then, the error is also increased, but remains below the $10^{-4}$ threshold for a thermal state $\bar n = 3$, even when accounting for the trap frequency inhomogeneity of $\pm 5~\%$ (gray zone). We also note that the error from the anharmonicity accumulates and becomes dominant for the next motion-echo time, at $\omega t_{\rm wait} = 1.5 \pi$, where it gets closer to our target fidelity. 
In conclusion, at this stage, the anharmonicity does not play an important role. However, we will see in the following sections that it becomes relevant when considering motion of the atom during the interaction pulse.

\section{Finite interaction time}
\label{sec:free-flight}

\begin{figure}
    \centering
    \includegraphics[width=\columnwidth]{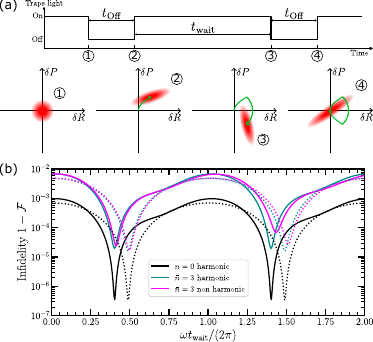}
    \caption{Motion-echo sequence adapted to finite interaction time. (a) Sequence timeline with phase-space representations at four steps. The first interaction step of duration $t_\text{Off}$ displaces and squeezes the initial state (step \textcircled{1} to \textcircled{2}). After trap evolution of duration $t_\text{wait}$ (step \textcircled{3}), a second free-expansion step restores the state to its initial position (step \textcircled{4}). (b) Gate infidelity as a function of the rotation angle $\omega t_\text{wait}$, for $t_\text{Off} = 10~\text{ns}$ (dashed) and $t_\text{Off} = 100~\text{ns}$ (full). The color code is the same as in Fig.\ref{fig:MotionEchoScheme}(d).}
    \label{fig:FreeSpaceFidelity}
\end{figure}

A key assumption in the previous section was the instantaneous acquisition of the interaction-induced phase. In practice, accumulating a $\pi/2$ phase shift would take on the order of tens of nanoseconds, limited by the finite van der Waals interaction strength, during which motional dynamics cannot be fully neglected. Instead of the straight displacement along the momentum axis, the phase-space trajectory of two interacting Rydberg atoms follows a curved trajectory depicted in Fig.\ref{fig:FreeSpaceFidelity}(a) as the atoms start to move away from each other (for a repulsive van der Waals potential). The motion-echo needs to account for this displacement. 

The situation is further complicated as the blue-detuned lattice confining the atoms in their ground state becomes much weaker once the atoms are excited to Rydberg states. For example, a state with principal quantum number $n = 60$ experiences a trapping depth approximately two orders of magnitude smaller. This would introduce motional dephasing due to the ground and Rydberg states evolving differently. Although this effect could in principle be mitigated using magic traps, we find it simpler to just switch off the trapping light during excitation such that both states experiences the same free-space expansion during the gate, leading to a state-independent squeezing of the wavefunction~\cite{Lienhard_2025}. An unsqueezing sequence can then coherently undo the free-space expansion at the end of the gate. 

Both effects are captured analytically in Appendix~\ref{app: Quadrature dynamics} via a Heisenberg transformation of the phase space quadratures. From this analysis, we derive new conditions for the motion-echo in Section~\ref{subsec: Freeflight kick cancelation}, and unsqueezing sequence to prevent heating during successive gate operations in Section~\ref{subsec: FreeFlight Squeezing cancel}.

\subsection{Momentum kick cancellation}
\label{subsec: Freeflight kick cancelation}

To cancel the phase-space displacement, we let the atoms oscillate in the trap by an angle $\omega t_\text{wait} = 2 \arctan(2/(\omega t_\text{Off})) \leq \pi $ (derived in Appendix~\ref{app: Quadrature dynamics}), such that the displacement induced during the second interaction step compensates that of the first, thereby closing the trajectory in phase space (Fig.~\ref{fig:FreeSpaceFidelity}(a) step \textcircled{3} and \textcircled{4}). 
As shown in Fig.\ref{fig:FreeSpaceFidelity}(b), for short interaction time (10 ns, dotted curves), the optimal echo angle is $\omega t_\text{wait} \simeq 0.98 \pi$, while it decreases to $0.80 \pi$ for the longer $t_\text{Off} = 100 ~\text{ns}$ (solid curves). 
This modified motion-echo sequence reaches similar performance as in the previous section when finite interaction was neglected, and we conclude that we can also succesfully disentangled internal and motional degrees-of-freedom in this more general and realistic case.

\begin{figure}
    \centering
    \includegraphics[width=\columnwidth]{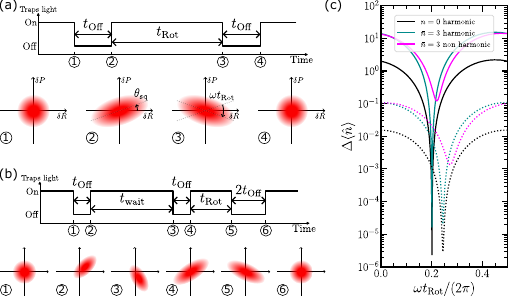}
    \caption{Heating-free gate sequence. (a) Squeeze-unsqueeze sequence. A first free-expansion step squeezes the initial state at angle $\theta_\text{sq}$ (step \textcircled{1} - \textcircled{2}). After a trap evolution of duration $t_\text{wait}$ (step \textcircled{3}), a second free expansion of equal duration restores the initial state (step \textcircled{4}). (b) Full sequence for the ultrafast gate under free-space expansion, combining motion echo and unsqueezing. Step \textcircled{1} - \textcircled{6} correspond to those in Fig.~\ref{fig:FreeSpaceFidelity}(a) and panel (a) with their associated phase-space representations. (c) Change in mean excitation number between initial and final states after the sequence described in (b), as a function of the rotation angle $\omega t_\text{Rot}$, for $t_\text{Off} = 10~\text{ns}$ (dashed) and $t_\text{Off} = 100~\text{ns}$ (full). The color code is the same as in Fig.\ref{fig:MotionEchoScheme}(d).}
    \label{fig:UnsqueezeSeq}
\end{figure}

\subsection{Residual squeezing mitigation}
\label{subsec: FreeFlight Squeezing cancel}

The adapted echo scheme restores the gate fidelity but significantly modifies motional state, as the free-space evolution squeezes it (Fig~\ref{fig:UnsqueezeSeq}(a), step \textcircled{2}). For a simple free-flight evolution without interaction, this resulting squeezed state is characterized by a squeezing factor $\beta( t_\text{Off})^2 = 1 + (\omega t_\text{Off})^2/2 + \omega t_\text{Off} \sqrt{1 + (\omega t_\text{Off})^2/4}$ and a squeezing angle $\theta_\text{sq}(t_\text{Off}) = \arctan \left( 2/(\omega t_\text{Off}) \right)/2$ 
For the echo sequence, which integrates two free-evolution steps combined with interaction, the final motional state is squeezed by a factor $\beta( 2t_\text{Off})$ at a non trivial angle $\theta_\text{sq}' = \arctan \left((4 + 3(\omega t_\text{Off})^2)/(\omega t_\text{Off})^3 \right)/2$ (Fig~\ref{fig:FreeSpaceFidelity}(a), step \textcircled{4}).

Due, for instance, to site-to-site variations in trapping frequencies, motional coherence is not preserved across successive computational steps. The gate is therefore effectively described as a heating process, increasing the mean excitation number by $\Delta \langle \hat n \rangle = (\omega t_\text{Off})^2 (2 \bar n + 1)$ ($\Delta \langle \hat n \rangle \sim 2.8$ for $t_\text{Off} = 100 ~\text{ns}$ and $ \bar n = 3$).

Each gate operation thus incrementally increases the mean motional excitation $\langle \hat n \rangle$, reducing the fidelity of subsequent operations. In a quantum computer requiring many high-fidelity gates, this effect sets a practical limit on the circuit depth before the atoms become too hot for reliable operation. Beyond this point, atoms must either be laser-cooled \cite{Bluvstein_2024, Bluvstein_2026}, or replaced by a freshly loaded ones \cite{Chiu_2025}. Nevertheless, minimizing gate induced heating remains crucial, as both cooling and reloading operations are time-consuming. We now show that a simple coherent motional control sequence can cancel the squeezing generated during the gate.

The unsqueezing protocol, shown in Fig.\ref{fig:UnsqueezeSeq}(a), consists of letting the state evolves in the trap for a duration $t_\text{Rot} = 2 \theta_\text{sq}(t_\text{Off})/\omega$, which flips the squeezed distribution in phase space ($\delta \hat R \rightarrow \delta \hat R$ and $\delta \hat P \rightarrow - \delta \hat P$). Subsequently, switching off the trapping potential for the same duration $t_\text{Off}$ restores the initial state.

This protocol is transposed to our gate in Fig.\ref{fig:UnsqueezeSeq}(b) by adding a final trap evolution time $ t_\text{Rot} = (\theta_\text{sq}'+ \theta_\text{sq}(2t_\text{Off}))/\omega$ (Fig.\ref{fig:UnsqueezeSeq}(b) step \textcircled{5}), followed by a free-space evolution of duration $2 t_\text{Off}$ (Fig.\ref{fig:UnsqueezeSeq}(b) step \textcircled{6}). As shown in Fig.\ref{fig:UnsqueezeSeq}(c), the mean excitation number is strongly suppressed at this optimal angle (namely $\omega t_\text{Rot} = 2\pi\times 0.202(0.245)$ for $100(10)~\text{ns}$ of interaction time). For short interaction times, the increase in the mean excitation number remains below 0.01 across all three dotted curves in Fig.~\ref{fig:UnsqueezeSeq}(c) over a range of rotation-angle fluctuations of $\pm 5 \%$. In contrast, for a 100 ns interaction time, the stronger squeezing narrows the range over which the sequence remains efficient at the $10^{-2}$ level. For a harmonic trap populated by a thermal state $\bar n = 3 $ (turquoise solid curve), the mean phonon number increase to $2.9 \times 10^{-2}$ when averaging over the full entangling zone (trapping frequency varying by $\pm 5 \%$). 

In this unsqueezing protocol, the anharmonicity plays a significant role: whether at 10 or 100 ns interaction time, it shifts the optimal angle to larger values (pink curves). This can be understood by noting that the squeezed atomic wavefunction explores a larger portion of the lattice potential, exploring more of the anharmonicity and slowing down the dynamics. Although this constrains the optimal squeezing removal operation, we still observe a reduction in the mean phonon by an order of magnitude, with an average over the entangling zone of $1.7 \times 10^{-3}$ ($1.5 \times 10^{-1}$) for 10 ns (100 ns), helping to suppress motional heating per gate operation.

\section{Entangling flux}
\label{sec:flux}
So far, we have shown that transporting atoms into a lattice interaction region and applying an echo sequence suppresses interatomic-distance uncertainty and spin-motion coupling. This strategy, however, imposes an architectural constraint --- entangling operations can be performed only within a designated gate zone, so atoms must be transported into and out of the lattice for each gate. It is therefore natural to ask what entangling throughput can be achieved in such an architecture. It is determined primarily by three quantities: the transport speed through the lattice, the number of atom pairs that can be addressed in parallel, and the minimum spacing between consecutively addressed rows.

If several gates are to be performed in parallel, the usable axial region is limited by the on-axis intensity variation of the two-dimensional lattice. Trap frequency tolerance of $10\%$ established in Sec.~\ref{motion-echo} restricts thus the operations to $L_{0.9}=400~\mu\rm m$ for the lattice parameters considered (see Sec.~\ref{sec:transverse_errors}), allowing simultaneous entanglement of approximately $40$ atomic pairs for an inter-pair spacing of $d=11~\mu\text{m}$, corresponding to a minimum interatomic separation of $8~\mu\text{m}$. At this separation, residual interactions between atoms belonging to different pairs contribute to phase shifts below $10^{-2}\pi$ corresponding to crosstalk errors below the $10^{-4}$ level for the geometry considered.

A final constraint arises from the requirement that only atomic pairs within the lattice interaction region be excited to the Rydberg states, while atoms outside this region remain unexcited. Consecutively addressed rows must therefore be separated sufficiently to suppress off-target Rydberg excitation. Assuming an excitation-beam waist comparable to that of the lattice beams, we take a representative row spacing of $l\simeq 2w_0=20~\mu\mathrm{m}$. The resulting aggregate entangling flux can then be estimated as $$\varphi\sim\frac{v_\text{max}}{l}\frac{L_{0.9}}{d}\approx 10^5~\text{gates}/\text{s}$$, where $v_\text{max}$ is the maximal transport speed (discussed in Section \ref{near_adiabatic_transport}), $d$ the inter-pair spacing, and $L_{0.9}$ the usable axial extent of the interaction region. We emphasize that this is an order-of-magnitude estimate, but indicates that gate rates at the $10^5~\text{s}^{-1}$ level are compatible with the suggested architecture. 

\section{Conclusion and outlook}

\begin{figure}
    \centering
    \includegraphics[width=\columnwidth]{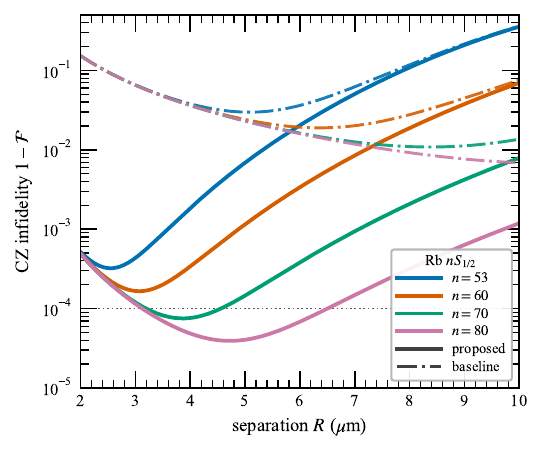}
    \caption{Total CZ-gate infidelity contribution from position noise and finite-lifetime $1-\mathcal{F}$ versus interatomic separation $R$ for $^{87}$Rb $nS_{1/2}$ pairs with $n=53,60,70,80$ (colors). Dash-dotted curves show the baseline interaction-driven gate in optical tweezers ($\omega_{x,y,(z),T}=2\pi\times100 (25)$~kHz); solid curves add the optical-lattice along \textit{x}, and \textit{z} and motion-echo protocols introduced here ($\omega_{x,L}=2\pi\times1$~MHz, $\omega_{z,L},~ \omega_{y,T}=2\pi\times100$~kHz). Each curve is the sum of the finite Rydberg-lifetime error and the position-dependent channels discussed in the text: interaction-phase miscalibration, first-order spin-motion coupling (replaced by its residual after the echo in the solid curves), van der Waals curvature, and transverse static and thermal motion. 
    Parameters: mean motional occupations $\bar n_{x,y,z}=3$, residual transverse offsets $\Delta x,~ \Delta y=50$~nm and $\Delta z=150$~nm, and an echo-timing spread of $\pm5\%$.}
    \label{fig:cz_infidelity}
\end{figure}

We have proposed a route towards interaction-driven Rydberg gates in which the dominant position sensitivity is suppressed by coherent motional control. Unlike blockade gates, whose speed is limited by the requirement $\Omega\ll V$, the interaction-driven gate accumulates its entangling phase directly from the Rydberg-Rydberg interaction, which lifts the blockade speed constraint, but makes the gate sensitive to the interatomic distance through $V\propto R^{-6}$. Our central result is that this sensitivity can be reduced with experimentally realistic tools. Figure~\ref{fig:cz_infidelity} summarizes the contribution to the CZ-gate infidelity by combining the error channels analyzed in this work with the finite Rydberg-state lifetime (with more details in the Appendix~\ref{app:error_budget}). Reducing the interatomic separation increases the interaction strength and therefore shortens the time spent in excited Rydberg states during the gate, suppressing lifetime-induced errors. At the same time, smaller separations amplify distance-induced phase errors and spin-motion coupling. Coherent motional control lowers the minimum achievable infidelity by two orders of magnitude, bringing the position noise gate-error contribution to the $10^{-4}$ regime. Two ingredients are responsible for this reduction.

First, the two-dimensional optical lattice defines a calibrated gate region in which the relevant length scale is set by optical periodicity rather than by tweezer placement alone. Along the interaction axis, the lattice suppresses tweezer-position errors by the factor $\omega_{x, T}^2/\omega_{x, L}^2$, converting $\sim100$~nm alignment errors into nanometer-scale distance errors. With accessible parameters, this reduces static phase errors to the $10^{-4}$ level while also tightening the motional wavepacket along the interaction axis. The second lattice dimension supplies additional confinement along the weakly trapped $z$ direction, suppressing the dominant residual thermal contribution once the longitudinal calibration error has been removed. We further find that atoms can be transported into and out of this lattice region with controlled motional heating excitation below $\Delta n\simeq10^{-2}$.

Second, a motion-echo sequence cancels the leading spin-motion coupling generated by the gradient of the van der Waals potential. Splitting the CZ gate into two $\sqrt{\rm CZ}$ interactions separated by half a trap period reverses the momentum kick from the first pulse and closes the phase-space trajectory. This removes the first-order sensitivity to quantum and thermal position fluctuations, with residual errors set by echo timing, transverse motion, van der Waals curvature, trap anharmonicity, and finite pulse duration. For the parameters considered here, these contributions remain compatible with errors below $10^{-4}$ for thermal occupations up to $\bar n\simeq3$.

We also showed that the protocol can be adapted to finite Rydberg-pulse durations, where the atoms undergo free expansion while the trapping light is switched off. By modifying the echo timing and adding an unsqueezing step, both the interaction-induced displacement and the associated motional heating can be suppressed, keeping the added excitation below $\Delta n\simeq10^{-2}$ for fast interaction times. Finally, the extended lattice gate zone allows many pairs to be addressed in parallel, giving an estimated entangling flux of order $10^5$ gates per second for representative parameters.

These results establish coherent control of atomic motion as a route to harnessing the full strength of Rydberg interaction to reach the two qubit gate infidelities at the $10^{-4}$ levels. The remaining challenge is to realize fast and selective excitation of a well-isolated pair state. If achieved, interaction-driven protocols would enable dynamics on timescales set directly by the interaction strength, suppressing errors from finite Rydberg lifetimes and other slow noise sources. 

During the completion of this work, we became aware of related study on \textit{positional-echo} applied to the entanglement of two molecules~\cite{Doyle2026}, as well as a proposal to engineer \textit{interaction-flatness} between Rydberg atoms of different species~\cite{PupilloSaffman2026}. 

\begin{acknowledgments}
W.A. thanks Matteo Simoni and Claudia Politi for feedback on the manuscript. 
SdL thanks Takuya Matsubara and Soonwoon Choi for discussion initiating this project at a workshop organized by the RIKEN Fundemental Quantum Science Program -  Harvard Quantum Innitiative. 
This work was supported by JST Moonshot R\&D Program Grant Number JPMJMS256D, and through JST-DFG 2024: Japanese-German
Joint Call for Proposals on “Quantum Technologies”
(Japan-JST-DFG-ASPIRE 2024) under JST Grand No.
JPMJAP24C2. 
\end{acknowledgments}

\appendix
\renewcommand\thefigure{S\arabic{figure}}   
\setcounter{figure}{0} 
\renewcommand{\thesection}{\Alph{section}}%
\setcounter{section}{0}

\nocite{*}

\section{Fidelity derivation under interaction phase errors}
\label{app: derivationSystPhaseError}

We derive the gate fidelity as a function of the phase shift acquired during the interaction step introduced in Sec.~\ref{sec: ultrafast gate}. The evolution operator after an interaction sequence of duration $t$ is
\begin{align}
    \hat U_{\rm Int}(t)
    &= \mathcal{P}_\perp + \exp\left[-\frac{i}{\hbar}\,V t\left(1+\frac{\Delta R}{R}\right)^{-6}\right] \mathcal{P}_{11}
\end{align}
where $\mathcal{P}_{11} = \ket{11}\bra{11}$, $\mathcal{P}_\perp = I - \mathcal{P}_{11}$.
The ideal target correspond to $Vt/\hbar = \pi$ with $\Delta R = 0$, which reduces to $ U_\text{ideal}$. Using the expression for the average gate fidelity, we obtain
\begin{equation}
    \mathcal{F} = \frac{ \left\lvert \text{tr}\left( U_\text{ideal}^\dagger \hat U_{\rm Int}(\pi\hbar/V) \right) \right \rvert^2 + d}{d(d+1)}
\end{equation}
with $d=4$ for a two-qubit that yields
\begin{equation}
    \mathcal{F} = \frac{7 - 3\cos(\phi)}{10}.
\end{equation}

\section{Two-dimensional lattice from tilted standing waves}
\label{app:2d_lattice}

As discussed in Sec.~\ref{sec:transverse_errors} and shown in Fig.\ref{fig:error_budget}, once the atoms are pinned by the one-dimensional lattice along $x$ and the residual spin-motion entanglement is removed by the echo sequence, the dominant remaining error arises from position fluctuations along the transverse direction $z$. We therefore consider replacing the one-dimensional lattice by a two-dimensional lattice that confines the atoms both along the interaction axis $x$ and along $z$. This can be implemented by replacing the single pair of counter-propagating beams along $x$ with two standing waves in the $x$-$z$ plane, tilted by angles $\pm\theta$ with respect to the $x$ axis. The resulting potential is sum of the two tilted standing-wave potentials. On the central axis, the ratio of the lattice trap frequencies is $\omega_{z,L}/\omega_{x,L}=\tan\theta$. Thus, transverse confinement can be introduced in a controlled way by choosing the tilt angle. The price is a reduced longitudinal region over which the lattice depth is approximately uniform. Increasing $\theta$ increases $\omega_{z,L}/\omega_{x,L}$, but also makes the atoms sample the transverse Gaussian envelope of the tilted beams more rapidly along the interaction axis. 

We estimate this usable length by asking over what range the local lattice frequencies remain at least $90\%$ of their central values. For a displacement $d$ along the central $x$-axis, and neglecting the longitudinal variation of the beam waist, valid for $d\cos\theta\ll z_R$, the local trap frequencies are reduced by the Gaussian envelope as

\begin{equation}
    \omega_{x(z),L}(d)
    =
    \omega_{x(z),L}(0)
    \exp\left[
        -\frac{d^2\sin^2\theta}{w_{0,L}^2}
    \right],
    \label{eq:tilted_lattice_frequency_envelope}
\end{equation}

where $w_0$ is the beam waist in the $x$-$z$ plane. The length over which the trap frequencies remain above $90\%$ of their central values is therefore

\begin{equation}
    L_{0.9}
    =
    2\frac{w_{0,L}}{\sin\theta}
    \sqrt{-\ln(0.9)}
    \simeq
    0.65\frac{w_{0,L}}{\sin\theta}.
    \label{eq:tilted_lattice_uniform_length_09}
\end{equation}

For the frequencies used in Sec.\ref{sec:transverse_errors}, $\omega_{x,L}=2\pi\times1{\rm MHz}$ and $\omega_{z,L}=2\pi\times0.1~{\rm MHz}$, one needs $\theta=\arctan(0.1)\simeq5.7^\circ$. This gives $L_{0.9}\simeq6.5w_{0,L}$. A circular beam with $w_{0,L}=10~\mu{\rm m}$ therefore provides a usable entangling region of only $L_{0.9}\simeq65~\mu{\rm m}$. 

A longer uniform region can be obtained with an elliptical lattice beam. Let $w_{xz,L}$ be the waist in the $x$-$z$ plane and $w_{y,L}$ the
orthogonal waist. We consider the regime $w_{xz,L}\gg w_{y,L}$, so that the Rayleigh range associated with the in-plane waist is much longer than that associated with the orthogonal waist, and so the longitudinal variation of $w_{xz,L}$ can therefore be neglected, while the Rayleigh divergence associated with $w_{y,L}$ must be retained. For a displacement $d$ along the central $x$ axis, each tilted beam is sampled a distance $d\cos\theta$ away from its waist along the beam propagation direction, and at a transverse offset $d\sin\theta$ in the $x$-$z$ plane. The resulting frequency envelope is
\begin{equation}
\begin{aligned}
    \omega_{x(z),L}(d)
    = \omega_{x(z),L}(0)
    &\left[
        1+
        \left(
            \frac{d\cos\theta\,\lambda}{\pi w_{y,L}^2}
        \right)^2
    \right]^{-1/4}\\ &
    \times\exp\left[
        -\frac{d^2\sin^2\theta}{w_{xz,L}^2}
    \right].
    \label{eq:tilted_lattice_frequency_envelope_elliptical}
\end{aligned}
\end{equation}

Setting Eq.~\eqref{eq:tilted_lattice_frequency_envelope_elliptical} equal to $0.9$ at the edge of the desired region, $d=L_{0.9}/2$, gives
\begin{equation}
    \frac{(L_{0.9}/2)^2\sin^2\theta}{w_{xz,L}^2}=-\ln(0.9)-
    \frac{1}{4}
    \ln\left[
        1+
        \left(
            \frac{L_{0.9}\cos\theta\,\lambda}{2\pi w_{y,L}^2}
        \right)^2
    \right].
    \label{eq:elliptical_waist_condition_with_y_rayleigh}
\end{equation}

Equivalently, the in-plane waist required to obtain a target usable length $L_{\rm target}$ is
\begin{equation}
    w_{xz,L}
    =
    \frac{(L_{\rm target}/2)\sin\theta}
    {
    \left(
    -\ln(0.9)
    -
    \frac{1}{4}
    \ln\left[
        1+
        \left(
            \frac{L_{\rm target}\cos\theta\,\lambda}{2\pi w_{y,L}^2}
        \right)^2
    \right]
    \right)^{1/2}
    } .
    \label{eq:required_elliptical_waist_with_y_rayleigh}
\end{equation}

For $L_{\rm target}=400~\mu{\rm m}$, $\theta\simeq5.7^\circ$, $w_{y,L}=10~\mu{\rm m}$, and $\lambda=775~{\rm nm}$, Equation~\eqref{eq:required_elliptical_waist_with_y_rayleigh} gives $w_{xz,L}\simeq88~\mu{\rm m}$. Thus, compared with a circular $10~\mu{\rm m}$ waist, the in-plane waist must be increased by a factor of about $8.8$. At fixed peak intensity, the required power per beam increases by the same factor for a single standing wave. In the two-standing-wave configuration of Sec.~\ref{sec:transverse_errors}, however, both lattices contribute to the longitudinal confinement, so each standing wave requires only half the depth and the per-beam power is reduced by a factor of two, yielding the $\approx140$~mW per beam quoted in the main text. This illustrates the basic tradeoff: stronger transverse confinement requires a larger tilt angle, while maintaining a long uniform entangling zone requires larger beam waists and optical power. Optimization of both waists can lead to decrease of the power required by a few percent.

\section{Diabatic excitation during transport}
\label{app:diabatic_transport}
\subsection{Effective one-dimensional potential}

To model transport through the lattice, we restrict the dynamics to a single axis and approximate both the lattice and tweezer potentials as harmonic. For the two-dimensional lattice, we choose the axis with the weaker tweezer confinement, since this direction sets the most stringent adiabaticity requirement. The coordinate $z$ denotes this weak motional axis, whose excitation we track. The time dependence enters through the Gaussian lattice envelope sampled as the tweezer is transported across the lattice beam in the orthogonal transverse direction, $y$. Thus, if the tweezer is initially displaced by $h$ from the lattice-beam center and is transported with velocity $v_0$, the sampled lattice potential is given by
\begin{align}
V_{L}^{(1\mathrm{D})}(z,t)
\simeq \frac{1}{2}m\,\xi(t)\omega_{z, L}^{2}z^{2},
\label{eq:Vl_expand}
\end{align}
with
\begin{equation}
\xi(t)=\exp[-2(h-v_0 t)^{2}/w_{y,L}^{2}],
\end{equation}
where $\omega_{z,L}$ is the lattice trap frequency at a site minimum, $w_{y, L}$ is the lattice beam waist in the y-direction.

Expanding the tweezer potential about its center $z_T$ gives
\begin{align}
V_{T}^{(1\mathrm{D})}(z)
\simeq \frac{1}{2}m\omega_{z,T}^{2}(z-z_T)^{2},
\label{eq:Vt_expand}
\end{align}
where $\omega_T$ is the tweezer trap frequency.

The total potential is therefore
\begin{align}
V_{\mathrm{tot}}^{(1\mathrm{D})}(z,t)
&=V_{L}^{(1\mathrm{D})}(z,t)+V_{T}^{(1\mathrm{D})}(z)
\nonumber\\
&\simeq \frac{1}{2}m\omega_{z,T}^{2}(z-z_T)^{2}
+\frac{1}{2}m\,\xi(t)\omega_{z,L}^{2}z^{2}
\nonumber\\
&\equiv \frac{1}{2}m\omega_z^{2}(t)\bigl[z-z_0(t)\bigr]^{2}
+\mathrm{const.},
\label{eq:Veff}
\end{align}
where the constant term only produces a global phase and is omitted. The effective trap frequency and instantaneous minimum are
\begin{equation}
\omega_z^{2}(t)=\omega_{z,T}^{2}+\xi(t)\omega_{z,L}^{2},
\qquad
z_0(t)=\frac{\omega_{z,T}^{2}}{\omega_z^{2}(t)}\,z_T .
\label{eq:omega_x0}
\end{equation}

\subsection{Diabatic excitation in a time-dependent harmonic trap}
The effective one-dimensional motional Hamiltonian corresponding to Eq.~\eqref{eq:Veff} reads
\begin{equation}
\hat H_z(t)=\frac{\hat p_z^{2}}{2m}+\frac{1}{2}m\omega_z^{2}(t)\big[\hat z-z_0(t)\big]^{2}.
\label{eq:H_eff}
\end{equation}
Introducing the instantaneous ground-state length
\begin{equation}
\delta_z(t)=\sqrt{\frac{\hbar}{2m\omega_z(t)}},
\end{equation}
we define the annihilation operator in the comoving frame as
\begin{equation}
\hat b_z(t)\equiv \frac{\hat z-z_0(t)}{2\delta_z(t)}+i\frac{\delta_z(t)}{\hbar}\hat p_z.
\label{eq:b_def}
\end{equation}
Because $\hat b_z(t)$ depends explicitly on time through $\omega(t)$ and $z_0(t)$, its Heisenberg-picture evolution is
\begin{align}
\frac{d}{dt}\hat b_{z,H}(t)
&=\frac{i}{\hbar}\big[\hat H_{z,H}(t),\hat b_{z,H}(t)\big]+\left(\frac{\partial \hat b_z}{\partial t}\right)_H
\nonumber\\
&=-i\omega_z(t)\hat b_{z,H}(t)+\frac{\dot\omega_z(t)}{2\omega_z(t)}\hat b_{z,H}^\dagger(t)-\frac{\dot z_0(t)}{2\delta_z(t)}.
\label{eq:b_eom}
\end{align}

In order to solve the dynamics, we remove the fast harmonic rotation by introducing the rotating-frame operator $\hat{c_z}(t)\equiv e^{i\phi_z(t)}\hat{b}_{z,H}(t)$, with
\begin{equation}
\phi_z(t)\equiv \int_{0}^{t}\omega_z(t')\,dt'.
\label{eq:phi}
\end{equation}

This allows us to transform Eq.~\eqref{eq:b_eom} into
\begin{align}
\frac{d}{dt}\hat c_z(t)
&=\frac{\dot\omega_z(t)}{2\omega_z(t)}e^{i2\phi_z(t)}\hat c_z^\dagger(t)-\frac{\dot z_0(t)}{2\delta_z(t)}e^{i\phi_z(t)}.
\label{eq:c_eom}
\end{align}

In the perturbative regime where $\dot\omega_z/\omega_z^2$ and $\dot z_0/(\omega_z\delta_z)$ are small on the timescale set by $\omega_z^{-1}$, we evaluate the right-hand side of Eq.~\eqref{eq:c_eom} using the zeroth-order solution $\hat c_z(t)\simeq \hat c_z(0)=\hat b_{z,H}(0)$. Integrating then yields

\begin{equation}
\begin{aligned}
\hat c_z(t) \approx\hat b_{z,H}(0) &+ \left(\int_0^t dt' \frac{\dot\omega_z(t')}{2\omega_z(t')}e^{i2\phi_z(t')}\right)\hat b^\dagger_{z,H}(0)\nonumber\\ &- \int_0^t dt' \frac{\dot z_0(t')}{2\delta_z(t')}e^{i\phi_z(t')}
\end{aligned}
\end{equation}

Transforming back to $\hat b_{z,H}(t)=e^{-i\phi_z(t)}\hat c_z(t)$ gives solution of the form

\begin{equation}
\hat b_{z,H}(t)\simeq u_z(t)\,\hat b_{z,H}(0)+\nu_z(t)\,\hat b_{z,H}^\dagger(0)+\beta_z(t),
\label{eq:affine_pert}
\end{equation}
with $u_z(t)\simeq e^{-i\phi_z(t)}$ and
\begin{align}
\nu_z(t) &\simeq e^{-i\phi_z(t)}\int_{0}^{t}\!dt'\,\frac{\dot\omega_z(t')}{2\omega_z(t')}\,e^{2i\phi_z(t')},
\label{eq:v_pert_phase}\\
\beta_z(t) &\simeq -e^{-i\phi_z(t)}\int_{0}^{t}\!dt'\,\frac{\dot z_0(t')}{2\delta_z(t')}\,e^{i\phi_z(t')}.
\label{eq:beta_pert_phase}
\end{align}
which correspond to Eq.\eqref{eq:v_approx} in the main text.

For an atom initially in the motional ground state of the initial trap, the excitation above the adiabatic ground state is therefore
\begin{equation}
\Delta n_z(t)\simeq |\nu_z(t)|^2+|\beta_z(t)|^2.
\end{equation}

For an initial thermal state with mean occupation $\bar n$, this generalizes to

\begin{equation}
\Delta n_z(t)\simeq (2\bar n_z+1)|\nu_z(t)|^2+|\beta_z(t)|^2.
\end{equation}

Near-adiabatic excitation along $x$-axis follows equivalent derivation.

\subsection{Specialization to a Gaussian lattice envelope}
Using Eq.~\eqref{eq:omega_x0} we obtain
\begin{equation}
\dot\omega_z(t)=\frac{\omega_{z,L}^{2}}{2\omega_z(t)}\,\dot\xi(t),
\qquad
\dot z_0(t)= -z_T\,\omega_{z,T}^{2}\omega_{z,L}^{2}\,\frac{\dot\xi(t)}{\omega_z^{4}(t)}.
\label{eq:omegadot_x0dot}
\end{equation}
Substituting Eq.~\eqref{eq:omegadot_x0dot} into Eq.~\eqref{eq:v_approx} yields
\begin{align}
\nu_z(t) &\simeq \int_{0}^{t}\!dt'\,
\frac{\omega_{z,L}^{2}\dot\xi(t')}{4\big[\omega_{z,T}^{2}+\xi(t')\omega_{z,L}^{2}\big]}\,
e^{- i\left[\phi_z(t)-2\phi_z(t')\right]},
\label{eq:v_xi}\\
\beta_z(t) &\simeq z_T\,\omega_{z,T}^{2}\omega_{z,L}^{2}\sqrt{\frac{m}{2\hbar}}
\int_{0}^{t}\!dt'\,
\frac{\dot\xi(t')}{\big[\omega_{z,T}^{2}+\xi(t')\omega_{z,L}^{2}\big]^{7/4}}\,\nonumber\\
&\hspace{3.6cm}\times
e^{-i\left[\phi_z(t)-\phi_z(t')\right]} .
\label{eq:beta_xi}
\end{align}

For $\xi(t)=\exp[-2(h-v_0 t)^{2}/w_{y,L}^{2}]$, Eqs.~\eqref{eq:v_xi}--\eqref{eq:beta_xi} are evaluated numerically in the main text using, e.g.,
$\omega_{z,L}=2\pi\times0.1~\mathrm{MHz}$, $\omega_{z,T}=2\pi\times25~\mathrm{kHz}$, $m=87~\mathrm{amu}$,
$w_{y,L}=10~\mu\mathrm{m}$, $h=40~\mu\mathrm{m}$ and $t=t_f =2h/v_0$.

\section{Numerical treatment of anharmonic transport}
\label{app:anharmonic_extensions}

To quantify motional heating during near-adiabatic transfer in the regime where the offsets $x_T$ and $z_T$ become comparable to the relevant tweezer length scales, it is necessary to account for the anharmonicity of the trapping potential, because the atom spends a significant fraction of the transport in the anharmonic part of the tweezer potential. For this purpose, we numerically propagate the motional wave function under the full time-dependent Hamiltonian, following a standard split-operator approach, as in Ref.~\cite{grochowski2025quantum}. 

Along the radial direction, the tweezer is the Gaussian profile displaced by $x_T$ from the origin,
\begin{equation}
    V_{x,T}^{(1D)}(x) = \frac{m\omega_{x,T}^2 w_T^2}{4}\Bigl(1 - e^{-2(x-x_T)^2/w_{x,T}^2}\Bigr),
\end{equation}
with tweezer waist $w_t$. The lattice is a retro-reflected standing wave with wavevector $k_L = 2\pi/\lambda_L$,
\begin{equation}
    V^{(1D)}_{x,L}(x,t) = \xi(t)\,\frac{m\omega_{x,L}^2}{2k_{x,L}^2}\,\sin^2\!\bigl(k_{x,L}x\bigr).
\end{equation}

Along the tweezer-beam propagation direction the relevant confinement is the axial Gaussian-beam profile, set by the Rayleigh range $z_R=\pi w_T^2/\lambda_{\rm tw}$ rather than the waist, displaced by $z_T$ from the origin,
\begin{equation}
    V_{z,T}^{(1D)}(z) = U_0\left(1 - \frac{1}{1+((z-z_T)/z_R)^2}\right),
\end{equation}
with depth $U_0$. In the axial geometry the lattice is generated by two beams tilted by $\pm\theta$ from the $z$-axis, so the standing wave along $z$ has the reduced effective wavevector $k_{\rm eff}=(2\pi/\lambda_L)\sin\theta$,
\begin{equation}
    V^{(1D)}_{z,L}(z,t) = \xi(t)\,\frac{m\omega_{z,L}^2}{2k_{z,L}^2}\,\sin^2\!\bigl(k_{z, L}z\bigr).
\end{equation}
The small tilt makes this lattice far coarser, and hence more harmonic across the wave packet, than a retro-reflected $\lambda_L/2$ lattice.

\subsection{Split-operator propagation}

The Hamiltonian is decomposed into kinetic and potential contributions, $\hat H=\hat T+\hat V^{(1D)}_L + \hat V^{(1D)}_T$, with $\hat T=\hat p^2/(2m)$. Since $\hat T$ is diagonal in momentum space and $\hat V$ is diagonal in position space, the short-time evolution operator can be implemented by trotterizing the unitary, using the symmetric splitting of the propagator.

\begin{equation}
    \hat U(dt)\approx
    e^{-i\hat V(t_{\rm mid})dt/(2\hbar)}
    e^{-i\hat T dt/\hbar}
    e^{-i\hat V(t_{\rm mid})dt/(2\hbar)},
\end{equation}
where $t_{\rm mid}=t+dt/2$. This approximation is accurate to $\mathcal O(dt^3)$ per time step.

Each propagation step is therefore carried out as follows. First, the wave function acquires a half-step phase in position space,
\begin{equation}
    \psi(x)\rightarrow
    e^{-iV(x,t_{\rm mid})dt/(2\hbar)}\psi(x).
\end{equation}
Next, the state is Fourier transformed to momentum space, where the kinetic evolution is applied,
\begin{equation}
    \psi(x)\rightarrow
    \mathscr{F}^{-1}\!\left[
    e^{-i\hbar k^2 dt/(2m)}
    \mathscr{F}[\psi](k)
    \right].
\end{equation}
Finally, a second half-step potential evolution is applied in position space.

\subsection{Extracting the mean excitation number}
After the full propagation, the residual excitation is quantified by projecting the final wave function onto the eigenstates $\{\ket{\phi_n}\}$ of the bare tweezer potential. 

\begin{equation}
    \bar{n} = \sum_{n=0}^{n_{\mathrm{eig}}-1} n\,|c_n|^2, \qquad
    c_n = \int\!\mathrm{d}x\;\phi_n^*(x)\,\psi_{\mathrm{final}}(x).
\end{equation}

The states $\ket{\phi_n}$ are obtained numerically by diagonalizing the static bare-tweezer Hamiltonian on the same spatial grid used for the time evolution, and we retain the lowest $n_{\mathrm{eig}}$ eigenstates in the projection. Evaluating $\bar n$ in a harmonic-oscillator basis would introduce an artificial excitation floor, because the physical tweezer is anharmonic.

\section{Near adiabatic transport along $x$-direction}
In Sec.~\ref{near_adiabatic_transport}, the weakly confined $z$-direction was identified as setting the dominant constraint on near-adiabatic transport. This follows from the larger tweezer trap frequency along $x$, which suppresses diabatic excitation for the same lattice envelope and transport speed. We verify this by repeating the one-dimensional numerical calculation for motion along the $x$ direction. This calculation is also relevant in case of using a one-dimensional lattice instead of two-dimensional one, where then the excitation along the $x$-direction constrains the maximal speed. 

\begin{figure}
    \centering
    \includegraphics[width=1\linewidth]{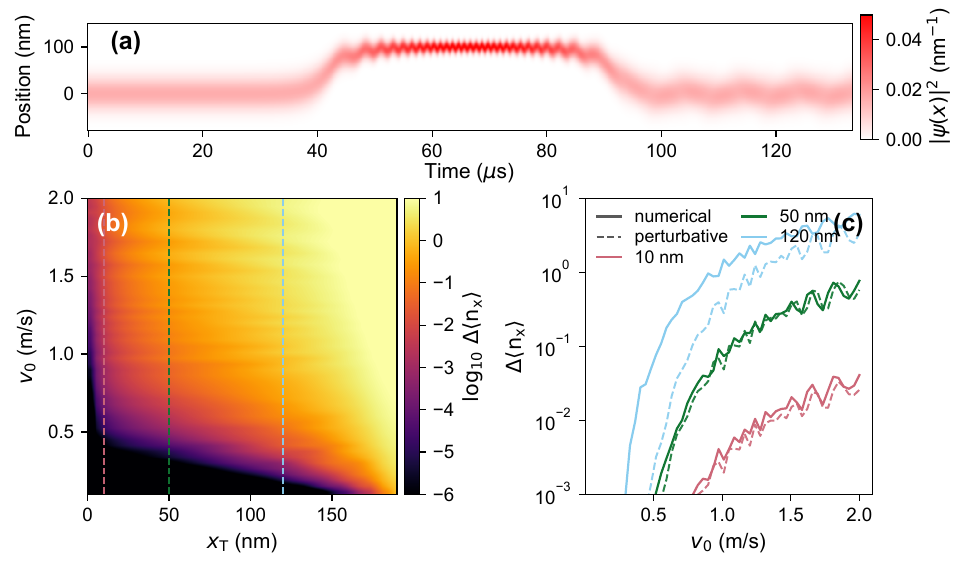}
    \caption{Transport-induced motional excitation in the lattice gate zone. (a) Probability density of an atom transported into and out of the lattice interaction region at $v_0=0.6~\mathrm{m/s}$ with tweezer–lattice offset $x_T=100~\mathrm{nm}$. The atom is pinned to the lattice minimum while traversing the interaction region; the residual oscillation after the passage reflects motional excitation. (b) Change in the mean vibrational occupation $\Delta\langle n_x\rangle$ after the full transport sequence, obtained from numerical propagation in the full time-dependent potential and averaged over a $10\%$ Gaussian spread in lattice trap frequency. (c) Numerical cuts from panel (b) compared with the perturbative prediction of Eq.~\eqref{eq:v_approx}. The harmonic theory accurately describes the small-offset regime, while deviations at larger $x_T$ mark the onset of anharmonic effects.}
    \label{fig: x-direction adiabatic transfer}
\end{figure}

The calculation follows the same procedure as in Appendix~\ref{app:anharmonic_extensions}. As in the main text, the result is averaged over a $10\%$ Gaussian spread in lattice trap frequency to remove narrow cancellation features that are not robust to site-to-site variations. Figure~\ref{fig: x-direction adiabatic transfer}(a) shows a representative trajectory for $v_0=0.6~\mathrm{m/s}$ and $x_T=100~\mathrm{nm}$. The lattice pins the atom during the passage, while the residual oscillation after leaving the lattice region gives the final motional excitation. Figures~\ref{fig: x-direction adiabatic transfer}(b,c) show that the $x$-direction excitation remains below the corresponding $z$-direction excitation over the relevant range of speeds and offsets. 

The main difference from the $z$-direction dynamics is the earlier onset of anharmonic corrections. The harmonic perturbative theory of Appendix~\ref{app:diabatic_transport} captures the small-offset regime, but deviations from the full numerical calculation appear already for $x_T\gtrsim100~\mathrm{nm}$. These deviations arise because, for a fixed absolute offset, the atom samples a larger fraction of the radial length scale of the Gaussian tweezer potential.

\section{Average gate fidelity with motional degrees of freedom}
\label{app: FidelyDer}
The computational basis is defined as $\mathcal{H}_{comp} = \{ \ket{00}, \ket{10}, \ket{01}, \ket{11}\}$. We additionally account for the dynamics of the motional state of the atoms in the trap, noting that only the state $\ket{11}$ undergoes interaction during its mapping to Rydberg states and subsequent evolution in non-harmonic potentials. 
The total system therefore evolves in the Hilbert space defined by the tensor product of the computational and motional subspaces, $\mathcal{H}_\text{comp} \otimes \mathcal{H}_\text{mot}$, where the latter is conveniently described in the Fock basis for the atomic motion. In this representation, the total evolution operator can be written as
\begin{equation}
    U_\text{total} = \mathcal{P}_\perp \otimes M_\perp - \mathcal{P}_{11} \otimes M_{11}
\end{equation}
where $\mathcal{P}_{11} = \ket{11}\bra{11}$, $\mathcal{P}_\perp = I - \mathcal{P}_{11}$, and $M_\perp$, $M_{11}$ denote the corresponding motional evolution operators.
We assume an initially separable state, $\rho_0 = \rho_\text{qbt} \otimes \rho_{mtn}$ and treat the motional degrees of freedom as an environment. Tracing over the environment after the gate, we obtain the reduced qubit channel of the form
\begin{align}
    \mathcal{E}(\rho_q) &= \text{tr}_{mtn}\left(U_\text{total} \rho_0 U_\text{total}^\dagger \right) \\
    &= \mathcal{K}_0 \rho_q \mathcal{K}_0^\dagger + \mathcal{K}_1 \rho_q \mathcal{K}_1^\dagger,
\end{align}
with Kraus operators $\mathcal{K}_0 = \mathcal{P}_\perp - s \mathcal{P}_{11}$, $\mathcal{K}_1 = \sqrt{1 - \vert s \vert^2} \mathcal{P}_{11}$, where $s = \text{tr}\left( M_{11} \rho_{mtn}  M_\perp^\dagger \right)$.

From this expression, the average gate fidelity can be evaluated as
\begin{align}
    \mathcal{F} = \frac{\sum_j \left\lvert \text{tr}\left( U_\text{ideal}^\dagger \mathcal{K}_j \right) \right \rvert^2 + d}{d(d+1)}
\end{align}
where the ideal evolution operator is $U_\text{ideal} = \mathcal{P}_\perp -  \mathcal{P}_{11}$ and $d = 4$ for a two-qubit computational space. This yields
\begin{align}
    \mathcal{F} = \frac{7 + 3\Re(s)}{10}
\end{align}

\section{Echo propagator and thermal-state overlap}
\label{app: DerivationEchoSequence}

In this section, we derive the motional propagator associated with the echo sequence introduced in Sec.~\ref{motion-echo}.
A quantitative description follows from the first-order expansion in Eq.~(\ref{eq:H_Int}). The motional evolution during a single interaction pulse of duration $t$ (conditioned on $\ket{\tilde{rr}}$ and acting on the relative motional Hilbert space) is
\begin{align}
    \hat U_{\rm Mot}(t)
    &= \exp\left[-\frac{i}{\hbar}\,V t\left(1-6\frac{\delta \hat R}{R}\right)\right]\nonumber\\
    &= e^{-i\phi}\exp\left[i\kappa(\hat a_{\rm rel}^\dagger+\hat a_{\rm rel})\right],
\end{align}
where $\delta\hat R=\delta R_0(\hat a_{\rm rel}^\dagger+\hat a_{\rm rel})$ defines the relative-mode ladder operators. Here we introduce $\kappa\equiv 6(Vt/\hbar)(\delta R_0/R)$ and $\phi\equiv Vt/\hbar$. The pulse is therefore a displacement in phase space of the relative mode,
\begin{equation}
    \hat U_{\rm Mot}=e^{-i\phi}\,\hat D_{\rm rel}(i\kappa).
\end{equation}
The intermediate free evolution for a wait time $t_{\rm wait}$ corresponds to a phase-space rotation
\begin{equation}
\hat R_{\rm rel}(\theta)=\exp\left(-i\theta\hat a_{\rm rel}^\dagger\hat a_{\rm rel}\right)
\end{equation}
where $\theta\equiv \omega t_{\rm wait}$ is the phase accumulated from wait time. The total motional propagator of the echo sequence is then
\begin{align}
\hat U_{\rm Mot}^{(\rm echo)}(\theta) &= \hat U_{\rm Mot}(t)\,\hat R_{\rm rel}(\theta)\,\hat U_{\rm Mot}(t) \nonumber\\
&= e^{-i2\phi}\, \hat D(i\kappa)\,\hat R_{\rm rel}(\theta)\,\hat D(i\kappa),
\end{align}
Using standard identity 
\begin{equation}
    \hat R(\theta)\,\hat D(\alpha) = \hat D(\alpha e^{-i\theta})\,\hat R(\theta),
\end{equation}
we commute the second displacement through the rotation to obtain 
\begin{equation}
    \hat U_{\rm Mot}^{(\rm echo)}(\theta) = e^{-i2\phi}\hat D_\text{rel}(i\kappa)\hat D_\text{rel}(i\kappa e^{-i\theta})\hat R_{\rm rel}(\theta).
\end{equation}
The problem reduces to considering two displacement operators. Using standard identity \begin{equation}
    \hat D(\alpha)\hat D(\beta) = \exp\left(\frac{\alpha\beta^*-\alpha^*\beta}{2}\right) \hat D(\alpha+\beta).
\end{equation}
Substituting $\alpha=i\kappa$ and $\beta=i\kappa e^{-i\theta}$ gives
\begin{equation}
    \hat D_\text{rel}(i\kappa)\hat D_\text{rel}(i\kappa e^{-i\theta}) = e^{\,i\kappa^2\sin\theta} \hat D_\text{rel}\left(i\kappa(1+e^{-i\theta})\right),
\end{equation}
which then gives the final result of 
\begin{equation}
    \hat U_{\rm Mot}^{(\rm echo)}(\theta) = e^{-i2\phi}\, e^{\,i\Phi_{\rm geo}(\theta)}\, \hat D_{\rm rel}\!\left(\alpha_{\rm res}(\theta)\right)\, \hat R_{\rm rel}(\theta)
\end{equation}
with residual displacement $\alpha_{\rm res}(\theta)=i\kappa\left(1+e^{-i\theta}\right)$ and geometric phase of $\Phi_{\rm geo}(\theta)=\kappa^2\sin\theta$.

Since the noninteracting branch undergoes the same free harmonic evolution (the overlap entering $s$ involves $M_{11}M_\perp^\dagger$) so the final rotation cancels. From this result, we obtain the expectation value $\langle e^{i2\phi} \hat U_{\rm Mot}^{(\rm echo)}(\theta) \rangle $ for a given Fock state $\ket n$, which reads
\begin{equation}
    \langle e^{i2\phi} \hat U_{\rm Mot}^{(\rm echo)}(\theta) \rangle = e^{-\vert \alpha_{\rm res}(\theta) \vert^2 /2} \mathcal{L}_n(\vert \alpha_{\rm res}(\theta) \vert^2),
\end{equation}
where $\mathcal{L}_n$ denotes the Laguerre polynomial. This result can be generalized to a thermal state $\rho_\text{th}$ with mean occupation number $\bar n$,
\begin{align}
    \langle e^{i2\phi} \hat U_{\rm Mot}^{(\rm echo)}(\theta) \rangle_{\bar n}  &= \text{tr}\left(\rho_\text{th}\hat D_{\rm rel}\!\left(\alpha_{\rm res}(\theta) \right) \right) \\
    & = \sum_k p_k e^{-\vert \alpha_{\rm res}(\theta) \vert^2 /2} \mathcal{L}_k(\vert \alpha_{\rm res}(\theta) \vert^2),
\end{align}
with $p_k = \bar n ^k/(\bar n +1)^{k+1}$. Using the generating function of Laguerre polynomials, we derive
\begin{equation}
    \langle e^{i2\phi} \hat U_{\rm Mot}^{(\rm echo)}(\theta) \rangle_{\bar n} = \exp \left(-\left(\bar n +\frac{1}{2} \right) \vert \alpha_{\rm res}(\theta) \vert^2 \right).
\end{equation}

\section{Transverse fluctuations errors derivation}
\label{app: Transverse fluctu}
In this section, we estimate the effect of the transverse extent of the motional wave function on the fidelity of the interaction-driven gate fidelity. We consider only position fluctuations along a single transverse axis (either $y$ or $z$). In this case, the interatomic distance can be written as $\hat R = R \sqrt{1 + (\delta \hat R_\perp /R)^2}$, where $\delta \hat R_\perp = \delta R_{\perp,0} \left( \hat a_\perp ^\dagger +  \hat a_\perp\right)$, which to leading order in $\delta \hat R_\perp /R$ becomes $\hat R \simeq R (1 + 1/2 (\delta \hat R_\perp /R)^2)$. The evolution operator corresponding to the accumulation of a $\pi$ phase can then be written and expanded to second order as
\begin{align}
    \hat U_{\rm Mot}
    &= \exp\left[-i \pi\left(1-3\left(\frac{\delta \hat R_\perp}{R} \right)^2\right)\right] \\
    & \simeq e^{-i \pi} \left[1 + 3 i\pi \left(\frac{\delta \hat R_\perp}{R} \right)^2 - \frac{9}{2} \pi^2 \left(\frac{\delta \hat R_\perp}{R} \right)^4\right].
\end{align}
The overlap $\Re (s)$ for a given Fock state $\ket{n}$ is written
\begin{align}
    \Re\left(\langle e^{i\pi} U_\text{Mot} \rangle \right) \simeq  1 - \frac{9}{2} \pi^2 \frac{\langle \delta \hat R_\perp^4 \rangle}{R^4},
\end{align}
with $\langle \delta \hat R_\perp^4 \rangle = 3 \delta R_{\perp,0}^4 \left(2 n^2 + 2n +1 \right)$. From this expression, the overlap can be evaluated for a thermal state with mean phonon number $\bar n$
\begin{align}
    \Re(s) &= \sum_k p_k \Re\left(\langle k \vert e^{i\pi} U_\text{Mot} \vert k \rangle \right)\\
    &\simeq 1 - \frac{3}{2} \left(3\pi\frac{\delta R_{\perp,0}^2}{R^2} \right)^2 (2 \bar n + 1)^2.
\end{align}

The effect of the van der Waals potential curvature on the fidelity can be evaluated using a similar derivation. We first consider the evolution operator obtained by retaining the second-order term in the expansion of the interaction potential
\begin{align}
    \hat U_{\rm Mot}
    &= \exp\left[-i \frac{\pi}{2} \left(1 - 6 \frac{\delta \hat R}{R} + 21 \left(\frac{\delta \hat R}{R} \right)^2\right)\right].
\end{align}
At the end of the motion-echo protocol, the total evolution operator is given by
\begin{align}
    \hat U_{\rm Mot}
    &= \exp\left[-i \pi \left(1 + 21 \left(\frac{\delta \hat R}{R} \right)^2\right)\right],
\end{align}
which leads to the following overlap for a thermal state :
\begin{equation}
    \Re(s) \simeq 1 - \frac{3}{2} \left(21 \pi\frac{\delta R_{0}^2}{R^2} \right)^2 (2 \bar n + 1)^2.
\end{equation}

\section{Quadrature description of finite-duration pulses}
\label{app: Quadrature dynamics}
In this section, we derive the Heisenberg evolution of normalized quadratures $\hat X = \delta \hat R/(\sqrt 2 \delta R_0)$ and $\hat P = \delta \hat P/(\sqrt 2 \delta P_0)$. We consider the general evolution operator $\hat{U} = \exp \left(-i a \hat{P}^2/2 + i b \hat X \right)$, which describes free-space evolution combined with a momentum kick. The action of this operator on the quadratures is obtained as follows,
\begin{align}
    \begin{pmatrix}
        \hat{U}^\dagger \hat{X} \hat{U}\\
        \hat{U}^\dagger \hat{P} \hat{U}
    \end{pmatrix} &=  \begin{pmatrix}
        1 & a \\ 0 & 1
    \end{pmatrix}  \begin{pmatrix}
        \hat{X}\\
        \hat{P}
    \end{pmatrix} + b\begin{pmatrix}
        \frac{a}{2} \\ 1
    \end{pmatrix}\\
    & = L \begin{pmatrix}
        \hat{X}\\
        \hat{P}
    \end{pmatrix} + M.
\end{align}
The same transformation is obtained for a trap evolution $\hat{U}(\theta) = \exp \left( -i \theta/2 \times\left(\hat{X}^2 + \hat{P}^2\right) \right)$ which is,
\begin{align}
    \begin{pmatrix}
        \hat{U}(\theta) ^\dagger \hat{X} \hat{U}(\theta)\\
        \hat{U}(\theta) ^\dagger \hat{P} \hat{U}(\theta)
    \end{pmatrix} & =
    \begin{pmatrix}
        \cos(\theta) & \sin(\theta) \\ -\sin(\theta) & \cos(\theta)
    \end{pmatrix}  \begin{pmatrix}
        \hat{X}\\
        \hat{P}
    \end{pmatrix} \\
    &= R(-\theta) \begin{pmatrix}
        \hat{X}\\
        \hat{P}
    \end{pmatrix}
\end{align}
In a motion-echo sequence, a succession of three operations $\hat U_\text{tot} = \hat{U} ~\hat{U}(\theta) ~\hat{U}$ is implemented, as described in the main text. The corresponding transformation of the quadratures can be expressed as a matrix product,
\begin{align}
    \begin{pmatrix}
        \hat U_\text{tot}^\dagger \hat{X} \hat U_\text{tot}\\
        \hat U_\text{tot}^\dagger \hat{P} \hat U_\text{tot}
    \end{pmatrix} &=   L \left[ R(-\theta) \left[ L \begin{pmatrix}
        \hat{X}\\
        \hat{P}
    \end{pmatrix} + M \right]\right]+ M\\
    &= L R(-\theta) L \begin{pmatrix}
        \hat{X}\\
        \hat{P}
    \end{pmatrix} + \left[L R(-\theta) + 1\right] M.
\end{align}
The constant term vanishes for $\theta = 2\arctan(2/a)$, corresponding to the optimal angle that cancels the interaction-induced kick. The remaining terms transform the quadratures through the successive application of two shearing matrices and a rotation, resulting in a squeezing of the initial state. The goal is then to determine the corresponding squeezing parameters in order to cancel this effect with additional steps.

First, the shearing matrix $L$ can be decomposed using singular value decomposition as
\begin{equation}
        L = R\left( \theta_2\right) \begin{pmatrix}
        \lambda & 0 \\
        0 & 1/\lambda
    \end{pmatrix} R\left( \theta_1 \right),
\end{equation}
where $\lambda^2 = 1 + a \left(a+\sqrt{a^2 + 4} \right)/2$, $\theta_1 = \arctan\left( 2/a \right)/2 - \pi/2$ and $\theta_2 = \arctan\left( 2/a \right)/2$. This decomposition shows that $L$ corresponds to a squeezing mapping in a rotated basis. Canceling this mapping is achieved by choosing $\theta = \arctan(2/a)$, such that
\begin{align}
    L R(-\theta) L &= R\left( \theta_2\right) \begin{pmatrix}
        \lambda & 0 \\
        0 & \frac{1}{\lambda}
    \end{pmatrix} R\left( - \frac{\pi}{2}\right) \begin{pmatrix}
        \lambda & 0 \\
        0 & \frac{1}{\lambda}
    \end{pmatrix} R\left( \theta_1 \right)\\
    &= R\left( \arctan \left(\frac{2}{a}\right) - \pi\right).
\end{align}

Secondly, to fully characterize the transformation $L R(-\theta) L$ for a given $\theta$, we adopt a similar approach by applying a singular value decomposition to
\begin{multline}
        \begin{pmatrix}
        \lambda & 0 \\
        0 & \frac{1}{\lambda}
    \end{pmatrix} R\left( \arctan \left(\frac{2}{a}\right) - \frac{\pi}{2} - \theta \right) \begin{pmatrix}
        \lambda & 0 \\
        0 & \frac{1}{\lambda}
    \end{pmatrix}.
\end{multline}
 We introduce $\theta' = \arctan \left(2/a\right) - \pi/2 - \theta$ and define $R'(\phi)$ as a reflection matrix around an axis at angle $\phi$, written as
\begin{equation}
    \begin{pmatrix}
        \cos(2 \phi) & \sin(2 \phi)\\
        \sin(2 \phi)& -\cos(2 \phi)
    \end{pmatrix},
\end{equation}
then we obtain
\begin{equation}
    \begin{pmatrix}
        \lambda & 0 \\
        0 & \frac{1}{\lambda}
    \end{pmatrix} R\left( \theta' \right) \begin{pmatrix}
        \lambda & 0 \\
        0 & \frac{1}{\lambda}
    \end{pmatrix}= R'(\theta_4) \begin{pmatrix}
        \mu & 0 \\
        0 & 1/\mu
    \end{pmatrix} R'(\theta_3),
\end{equation}
where $\mu^2 = 1 + \gamma \left(\gamma + \sqrt{\gamma^2 + 4} \right)/2$, with $\gamma = \vert \cos(\theta')\vert a \sqrt{a^2 + 4}$, $\theta_3 = -\arctan(2\tan(\theta')/(2+a^2))/4$ and $\theta_4 = \pi/2 - \theta_3$. For the kick-canceling angle $\theta = 2\arctan(2/a)$, one finds $\gamma = 2 a$, corresponding to the same squeezing strength as $L^2$. The associated squeezing angle is then $\arctan((4 + 3 a^2)/a^3)/2$. 

\section{Motional error budget}
\label{app:error_budget}

\begin{figure*}
    \centering
    \includegraphics[width=\textwidth]{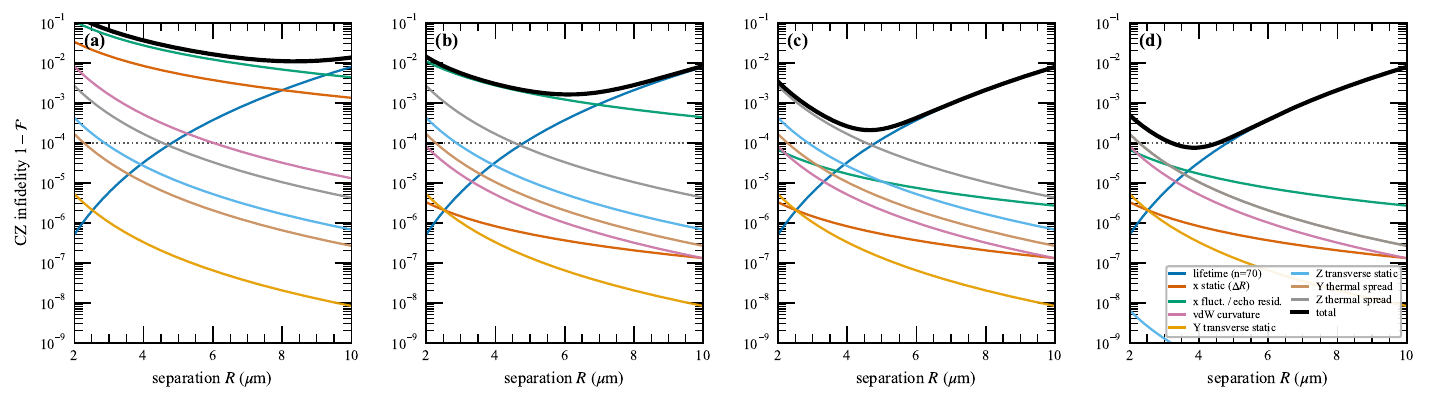}
    \caption{Error budget of the interaction-driven CZ gate versus interatomic separation $R$, for $^{87}$Rb $70S_{1/2}$. Each panel decomposes the total infidelity (thick black) into its individual contributions for one configuration: (a) baseline interaction gate, (b) interaction gate in one-dimensional optical lattice along x, (c) one dimensional lattice along x plus the motion-echo sequence, and (d) lattices along x, and z plus the motion-echo sequence. Colored curves show error contributions from: finite Rydberg lifetime, interaction-phase miscalibration ($x$-static, $\Delta R$), first-order spin-motion coupling along $x$ [the un-echoed fluctuation in (a,b), its echo residual in (c, d)], van der Waals curvature, and transverse ($y,z$) static and thermal motion. The transverse contributions are identical in (a, b, c) panels because the one-dimensional lattice provides no transverse confinement and the echo is timed to $\omega_{x,L}$. In (d) we introduce additional lattice along z, which decreases the error for the z static and thermal contribution. The dotted line marks $10^{-4}$. Parameters as in Fig.~\ref{fig:cz_infidelity}: $\omega_{x, L}=2\pi\times1$~MHz, $\omega_{z,L}=2\pi\times100$~kHz, $\omega_{x,T}=\omega_{y,T}=2\pi\times100$~kHz, $\omega_{z,T}=2\pi\times25$~kHz, $\bar n_{x,y, z}=3$, $\Delta x=\Delta y=50$~nm, $\Delta z=150$~nm, and an echo-timing spread of $\pm5\%$.}
    \label{fig:error_budget}
\end{figure*}

The interaction-driven CZ infidelity reported in the main text (Fig.~\ref{fig:cz_infidelity}) is obtained by summing the independent error channels analyzed above, each evaluated at leading order in its small parameter. Figure~\ref{fig:error_budget} decomposes this budget as a function of the interatomic separation $R$ for a representative $^{87}$Rb $70S_{1/2}$ pair. We consider here four configurations of increasing complexity (baseline interaction gate, interaction gate in a one dimensional optical lattice along x, interaction gate in a one dimensional lattice along x with motion echo, and interaction gate in lattices along x and z with motion echo) in order to capture the individual contribution of proposed protocols.

The contributions are:

\textit{(i) Finite Rydberg lifetime.} Both atoms may decay during the gate time $t_{\rm CZ}$, giving $\epsilon_{\rm life}=1-\exp(-2t_{\rm CZ}/\tau)$. We use the $C_6$ coefficients and $300$~K lifetimes (including blackbody-induced decay) from ARC~\cite{ARC}; for $70S_{1/2}$.

\textit{(ii) Interaction-phase miscalibration ($x$-static).} A static distance error $\Delta R$ gives rise to gate error given by Eq.~\ref{eq:fidelity_phase_uncertainty}. In the baseline interaction gate positioning error is $\Delta R=\Delta x=50$~nm; in the lattice configurations the atoms are pinned to the lattice minima and this error is suppressed to $\Delta R\approx0.5$~nm.

\textit{(iii) First-order spin-motion coupling ($x$).} Quantum and thermal fluctuations of the relative coordinate. Without the echo [panels (a,b)] this gives
$\tfrac{3}{20}(6\pi\,\delta R_0/R)^2(2\bar n_x+1)$; the lattice reduces $\delta R_0$ from $34$~nm ($\omega_{x,T}=2\pi\times100$~kHz) to $11$~nm ($\omega_{x,L}=2\pi\times1$~MHz). The motion-echo cancels this first-order term [panel (c, d)], leaving only the timing residual error given by Eq.~\ref{eq:residual_echo_error} with an echo-timing spread of$\Delta\theta=\pm5\%$.

\textit{(iv) van der Waals curvature.} The second-order term of the interaction expansion is not removed by the echo and contributes with an error given by Eq.~\ref{eq:residual_van_der_valse}, present in all four panels.

\textit{(v) Transverse static offsets.} A static transverse distance error $\Delta_\perp$ shifts the separation only at second order, $\Delta R\simeq\Delta_\perp^2/2R$, yielding $\tfrac{3}{20}(3\pi\,\Delta_\perp^2/R^2)^2$ for each of $y$ and $z$, with $\Delta y=50$~nm and $\Delta z=150$~nm. Introduction of the lattice along z reduces the z-positioning error to $9$~nm [panel (d)].

\textit{(vi) Transverse thermal motion.} Transverse fluctuations contribute with an error given by Eq.~\ref{eq:transverse_positioning_error_fluctuation} for each axis, with ground-state extents $\delta_{y,0}=34$~nm ($\omega_y=2\pi\times100$~kHz) and
$\delta_{z,0}=68$~nm ($\omega_{z,T}=2\pi\times25$~kHz), and occupations $\bar n_y=\bar n_z=3$. Introduction of the lattice along z reduces the z ground state extent to $34$~nm.

\bibliography{ultrafast_gate.bib}

\end{document}